\documentclass[twocolumn,showpacs,preprintnumbers,amsmath,amssymb,prb,floatfix,superscriptaddress]{revtex4}

\usepackage{graphicx,tabularx}
\usepackage[version=3]{mhchem} % Formula subscripts using \ce{}
\usepackage{amssymb}
\usepackage{dcolumn}
\usepackage{color}% to create grayscales for the ball and stick images
\usepackage[mathcal]{euscript}
\usepackage{color}
\definecolor{gray0}{gray}{0.0}%black
\definecolor{gray64}{gray}{0.25}
\definecolor{gray128}{gray}{0.5}
\definecolor{gray192}{gray}{0.75}
\definecolor{gray255}{gray}{1.0}%white

\begin{document}

\title{Modeling 1D structures on semiconductor surfaces: Synergy of theory and experiment.}
\author{Danny E. P. Vanpoucke}
\affiliation{Center for Molecular Modeling, Ghent University, Technologiepark $903$, $9053$ Zwijnaarde, Belgium}

\date{\today}
\begin{abstract}
Atomic scale nanowires attract enormous interest in a wide range of fields. On the one hand, due to their quasi-one-dimensional nature, they can act as a experimental testbed for exotic physics: Peierls instability, charge density waves, and Luttinger liquid behavior. On the other hand, due to their small size, they are of interest for future device applications in the micro-electronics industry, but also for applications regarding molecular electronics. This versatile nature makes them interesting systems to produce and study, but their size and growth conditions push both experimental production and theoretical modeling to their limits. In this review, modeling of atomic scale nanowires on semiconductor surfaces is discussed focusing on the interplay between theory and experiment. The current state of modeling efforts on Pt- and Au-induced nanowires on Ge(001) is presented, indicating their similarities and differences. Recently discovered nanowire systems (Ir, Co, Sr) on the Ge(001) surface are also touched upon. The importance of scanning tunneling microscopy as a tool for direct comparison of theoretical and experimental data is shown, as is the power of density functional theory as an atomistic simulation approach. It becomes clear that complementary strengths of theoretical and experimental investigations are required for successful modeling of the atomistic nanowires, due to their complexity.
\end{abstract}

%\pacs{  } %--> wtf???
%----------------------------------------------------------------
\maketitle
%------------------------------------------------------------------------------------------
%------------------------------Introduction------------------------------------------------
%------------------------------------------------------------------------------------------
\tableofcontents %generates TOC
%and a quickfix:
\makeatletter
\let\toc@pre\relax
\let\toc@post\relax
\makeatother

\section{Introduction}
\indent Everybody knows Moore's Law, \cite{Moore:elec65} or at least has a vague idea of its consequences: ``Next years computer will be faster.'' In $1965$, Gordon Moore observed that the number of components per integrated circuit, that could be produced at the lowest cost, doubled oughly every year.\cite{fn:Moore} Meanwhile, this primarily economical `law' has meanwhile become a self-fulfilling prophecy, driving the micro-electronics industry. Current fourth generation Intel Core chips are based on $22$ nm technology, and $4$ nm technology is expected to be introduced in commercial end-user applications around $2022$.\cite{Xbit_4nm} However this miniaturization cannot be maintained indefinitely and modern lithographical techniques are expected to meet their limits in the current decade.\cite{webpage:NewLifeMoore} Moreover, miniaturization is also steadily approaching its ultimate and final limit: atomic size devices connected by atomic wires.\cite{Kurzweil} To build these ultimate devices on an industrial scale, chip makers are looking toward self-assembly of surface nanostructures and nanowires (NWs).\cite{Barth:nat05} Approaching the atomic scale region, quantum effects become increasingly important with regard to the behavior and operation of these nanoscale devices. As a result, atomic scale modeling at the quantum mechanical level becomes an essential tool for understanding and designing such devices.\\
\indent Besides practical applications, self-assembled atomic scale NWs are also of interest from the fundamental point of view. Due to their inherent one-dimensional ($1$D) nature, they provide interesting model systems to study the physics of low dimensional systems and the associated exotic phenomena. As such, NWs provide ideal systems for studying dimensionality effects on, for example, the electronic structure and magnetism of a material.\cite{Shen:prb97, DoranD:prl98, Gamb:nat02, Nilius:sc02, Crain:sc05, Lagoute:prb06, LimDK:nano07, Hong:prb07} Atomic scale NWs also provide the opportunity to critically test predictions of solid state physics, such as the \textit{Peierls instability}: a metal-insulator transition introduced by the presence of a \textit{charge density wave} (CDW).\cite{YeomHW:PhysRevB2002, GuoJiandong:PhysRevLett2005, SnijdersPC:PhysRevLett2006, Houselt:ss08, ZandvlietHJW:JPhysCondMatter2009, SnijdersPC:RevModPhys2010, MochizukiI:PhysRevB2012} Furthermore, in $1$D electron systems, which could be present in some of these NW systems, the Fermi-liquid approach is predicted to break down.\cite{VoitJ:RepProgPhys1995} In such systems, \textit{Tomonaga-Luttinger theory} would describe the properties in a much better fashion, making self-assembled NWs an ideal testbed.\cite{Lutt:jmp63, KaneCL:PhysRevLett1992, SegoviaP:Nature1999, LosioR:PhysRevLett2001, ThielemannB:PhysRevLett2009, NakatsujiK:natphys2012, BlumensteinC:natphys2012Reply}\\
\indent $1$D structures have been grown on both metallic and semi-conductor surfaces, resulting in a large variety of different systems. This variety in systems in turn has led to a broad spectrum in terminology: chains, (nano)wires, nanolines, stripes, rods, \textit{etc.} Definitions differ from author to author and overlap between terms exists. For simplicity, we will use the term ``nanowire'' to refer to all these structures in general.\\
\indent Si surfaces have received a significant amount of attention, due to their importance in the semi-conductor industry. Indium atoms on Si(111) form NWs with a metallic character,\cite{Yeom:prl99, GonzalezC:PhysRevLett2006, AhnJR:PhysRevB2007, StekolnikovAA:PhysRevLett2007, WippermannS:PhysRevLett2008, GonzalezC:PhysRevLett2009, ChandolaS:PhysRevLett2009, HattaS:PhysRevB2011} but also on high-index surfaces, such as Si(557), atomic wires are observed.\cite{ShinBG:ApplPhysLett2013} This is similar to Au, which preferentially forms chains on the high-index surfaces like Si(553)\cite{Ahn:prl05, SnijdersPC:PhysRevLett2006, CrainJN:ApplPhysA2006, SnijdersPC:RevModPhys2010, ShinJS:PhysRevB2012} and Si(557)\cite{SegoviaP:Nature1999, CrainJN:ApplPhysA2006, RiikonenS:PhysRevB2007, HanJH:PhysRevB2009, KrawiecM:PhysRevB2010, SnijdersPC:RevModPhys2010, HoganC:PhysStatSolB2012, HoganC:PhysRevLett2013}. Also Pb on Si(557) gives rise to conducting NWs, with quasi-1D states below a critical temperature of $78$K; above this critical temperature the two dimensional (2D) coupling of the NWs makes the Pb chains 2D conducting.\cite{TegenkampC:PhysRevLett2005, TegenkampC:PhysRevLett2008, BlockT:PhysRevB2011}\\
\indent A different kind of NWs are the Bi nanolines on Si(001). These reconstructions are not metallic, but one of the few examples of atomically perfect 1D systems on Si.\cite{NaitohM:SurfSci1997, MikiK:PhysRevB1999, NaitohM:SurfSci2001, OwenJHG:PhysRevLett2002} A good review on these structures is found in Ref.~\onlinecite{OwenJHG:JMaterSci2006}. The metallic rare-earth silicides form another family of $1$D systems, with phenomenological similarities to the above Bi nanolines; They also received significant attention over the years.\cite{PreinesbergerC:JPhysDApplPhys1998, ChenY:ApplPhysLett2000, PreinesbergerC:JApplPhys2002, OwenJHG:JMaterSci2006, WankeM:PhysRevB2011} In these systems, the rare-earths give rise to 1D structures due to the anisotropy in the heteroepitaxial strain between the Si(001) surface and the rare-earth silicide, \textit{i.e.} the rare-earths can obtain a close lattice match with Si(001) in one direction while a large lattice mismatch exists in the orthogonal direction.\cite{PreinesbergerC:JPhysDApplPhys1998} The resulting 1D structures have widths in the range of $3$--$10$ nm, while their lengths can extend for hundreds of nanometers.\\
\indent As becomes clear from the above, the list of possible materials to deposit on Si, which result in $1$D structures has grown considerably over the last decade. A quick survey of the literature provides us with numerous examples: Mg/Si(557)\cite{ShinBG:SurfSci2012}, Si/Si(001)\cite{BiancoF:PhysRevB2011}, Mn/Si(001)\cite{LiuH:SurfSCi2008, WangJT:PhysRevLett2010, NolphCA:SurfSci2011, LiuHJ:JPhysCondMat2012, SimovKR:JPhysChemC2012}, \ce{CoSi2}/Si\cite{HeZ:PhysRevLett2004}, Ga/Si(001)\cite{WangJZ:PhysRevB2002, AlbaoMA:PhysRevB2005, PavelK:PhysRevB2006_Comment, AlbaoMA:PhysRevB2006_REPLY}, Sr/Si(111)\cite{ZhachukRA:PhysSolState_2010} and Sr/Si(001)\cite{ReinerJW:AdvMater2010}, Ag+Au/Si(557),\cite{KrawiecM:PhysRevB2010} Ir/Si(001)\cite{OncelN:JPhysCondMat2013}, and so on.\\
\indent In the quest for ever faster and smaller electronics, Ge is considered one of the most promising alternative materials for Si, since its lower effective hole mass and higher electron and hole drift mobility allows for higher switching speeds.\cite{SzeSM:SolidStateElectron1968, SaraswatKC:MicroelectronEng2005, BrachtH:PhysRevLett2009} In addition to high speed microelectronics, metal/Ge systems are also of importance in the development of highly sensitive radiation detection systems and in recent decades NW structures have been observed for several metal/Ge systems. Pelletier \textit{et al.}\cite{Pelletier:JVS2000} noted the formation of NWs when depositing Er on Ge(111) at $300^{\circ}$C. Other rare-earth nanorods are reported for example by Eames \textit{et al.}\cite{Eames:prb06} after deposition of Ho. The In/Ge(001) system has a somewhat longer history.\cite{Rich:prb1990, GaiZheng:SurfSci1995, Seehofer:ss1996, Falkenberg:ss1997, Falkenberg:prb2002, CakmakM:SurfSci2004} Submonolayer deposition of In on Ge(001) shows a complex reconstruction behavior, with models of In chains on Ge(001) already present in $1990$ in the work of Rich \textit{et al.}\cite{Rich:prb1990} in $1990$. Later work by Falkenberg \textit{et al.} \cite{Falkenberg:prb2002} proposes rather complex reconstructions for the observed In induced NWs. More recently Pt and Au NWs on Ge(001) attracted attention.\cite{SchaferJ:NewJPhys2009} After deposition of (sub)monolayer amounts of Pt\cite{Gurlu:apl03, Gurlu:prb04, Oncel:prl05, Oncel:ss06, Schafer:prb06, Fischer:prb07, Houselt:nanol06, Houselt:ss08, Kockmann:prb08, ZandvlietHJW:JPhysCondMatter2009, MochizukiI:PhysRevB2012, HeimbuchR:JPhysCondensMatter2013} or Au\cite{WangJ:PhysRevB2004, WangJ:SurfSci2005, Schafer:prl2008, vanHouseltA:JPhysCondensMatter2010, MockingTF:SurfSci2010, MeyerS:PhysRevB2011, NiikuraR:PhysRevB2011, MeyerS:PhysRevB2012, MelnikS:SurfSci2012, BlumensteinC:JPhysCondensMatter2013} on Ge(001), large NW arrays were observed by several groups. These NWs have a width of the order of a single atom and their length is only limited by the underlying plateau, resulting in huge aspect ratios.\cite{Gurlu:apl03, Schafer:prb06, WangJ:PhysRevB2004, Schafer:prl2008} Note that the width of these NWs is at least one order of magnitude smaller than that of the rare-earth induced NWs on Si, making them much more suited to test 1D electron behavior. The possibility of decorating the Pt NWs with CO molecules makes them interesting structures to study (and build) molecular electronics, but also to test electronic transport through a single (octanethiol) molecule.\cite{Oncel:ss06, Kockmann:prb08, HeimbuchR:PhysRevB2012, KumarA:JPhysCondensMatter2012}\\
\indent Recently, also Co deposited on Ge(001) was shown to give rise to 1D structures, as expected from the Co-Ge phase diagram,\cite{VanpouckeDannyEP:2009ThesisNW} albeit with slightly larger building blocks consisting of hexagonal structures containing Co atoms.\cite{ZandvlietHJW:SurfSci2011} Another recent example are Ir NWs on Ge(001), which were studied by Mocking \textit{et al.}\cite{MockingTF:naturecomm2013} and were suggested to present standing wave patterns in the NWs due to conduction electrons scattering at the ends of the NW.\\
\indent It is clear that there is a large number of materials that give rise to the formation of 1D structures on semiconductor surfaces, many of which have been observed already; still more are awaiting discovery. Despite the structural and chemical similarity between Si and Ge, the same metals that gives rise to NWs on Si, does not always give rise to NWs on Ge and vice versa.\cite{Kageshima:ss01, ItohH:PhysRevB1992} As a result, a model for each system needs to be developed from scratch.\\
\ \\
\indent With the miniaturization-drive pushing into the atomic scale sizes, theoretical atomic scale modeling becomes a relevant tool for practical real-life devices, since this is the only way to fully understand and predict their quantum mechanical behavior. Furthermore, the resulting experimental systems have building blocks with sizes that allow for direct comparison to theoretical models, without the need for making additional assumptions on the effects of scaling, or the indirect measurement of atomic properties.\cite{WenmackersS:StatNeer2012}\\
\indent This direct comparison is largely indebted to two developments. First, the development of the Scanning Tunneling Microscope (STM) has allowed systems to be studied at atomic length scales by experimentalists, and second, the advances in electronic structure modeling, in particular Density Functional Theory (DFT), allow theoreticians to model atomic structures up to a few hundred atoms routinely. Both methods have their limitations: STM cannot show which atomic species is located at which position, while DFT does not account for temperature and pressure. Therefor, close collaboration between the experimental and theoretical side is an essential part in the successful study of atomic scale structures.\\
\indent In this review, metal/Ge(001) atomic scale NWs will be discussed, with an emphasis on Pt and Au NWs. The benefits and necessity of direct comparison between experiment and theory for the model development will be brought to attention. Each system will be presented as a separate section. Since a good theoretical model should both describe existing experimental data, and predict future experiments, each of these sections starts with a subsection giving an overview of the experimental background, with a focus on structural information available from STM experiments. A second subsection presents the models available in the literature, and compares these to the experimental data, indicating strengths and weaknesses. In the final subsection aspects that need further investigation are indicated. We end this review with an outlook on metal-induced NWs, indicate some materials that may be interesting candidates for NW formation and present some aspects which are of importance for NW modeling and design.\\
\indent Because of their important role in the development of high quality models for atomic scale NWs, we start with a short summary of STM and DFT in section~\ref{Methods}.

%Such a temperature dependent transition in the geometric and/or electronic structure of a nanowire can also be an indication of the presence of a Peierls instability, as is suggested for ,

%However, the formation of robust ''one''-dimensional nanowires with a width less than  10 nm has been a major goal that has proven difficult to  achieve by either epitaxial growth or lithographic processing.1

%\indent In this thesis, the focus will go mainly to: Pt nanowires on Ge(001), as they are referred to in experimental literature. More specifically, I will try to elucidate its geometrical and electronic structure and relate this to observed experimental features. Since most of this work uses direct comparison with experiment, a description of the experiments and their results should not be omitted. In fact, it is the starting point of this story.

\section{Methods}\label{Methods}
\indent DFT and STM are of great importance during the study of surfaces and their reconstructions. The progress in the development of DFT currently allows the investigation of systems up to a few hundred atoms in size at the quantum mechanical level. Combined with periodic boundary conditions in two or three dimensions, this allows one to study surfaces as single unit cells. Since DFT is a ground-state theory, it is perfectly suited to identify the surface reconstruction with the lowest energy (from a set of investigated systems). On the other hand, STM shows the actual structure of a surface (reconstruction) at the atomic scale of a real sample. Although both methods appear to provide the same information, they do not. STM has no chemical insensitivity: for a material with multiple types of atoms, it is unclear which atom is seen at which position. Moreover, STM does not actually show ``atoms'' (as it is often portrayed); it shows atomic bonds and dangling bonds. DFT on the other hand does have the chemical sensitivity, but it is limited by the fact that the user needs to provide the crystal structure (or at least a starting configuration close enough to the actual structure). If the actual experimental structure is not present in the set of investigated structures, one will end up with the wrong model for the experimental structure. This limitation is not present in STM, since nature always provides it with the correct atomic structure, although we can observe it only indirectly. As such, DFT and STM can be seen as complementary methods: only by combining results from both it is possible to obtain a full picture of the experimental structure under study. In this review, we will see that without this synergy, finding a good model of surface reconstructions---NWs in this case---becomes nearly impossible. In the following, we present the two techniques in more detail.\\
\subsection{Electronic structure calculations: Density Functional Theory}
\indent In quantum mechanics, a system is fully described by its Schr\"{o}dinger equation. However, already for the general two-body problem no analytic solution exists. In contrast, numerically exact solutions can be obtained for many-body systems, although these tend to become computationally intractable quickly. Soon, it became clear that some level of approximation is needed to be able to model such systems. Already in $1928$, Hartree proposed to approximate the $N$-body wavefunction as a product of $N$ single particle wavefunctions. In this approximation, the interactions, between an electron and all other electrons in the system, are replaced by a single interaction with an averaged potential due to the other electrons. Hence, this approximation is called a mean-field approximation. The exchange contribution is an important contribution to the electron-electron interaction, which is missing in the Hartree approximation, as was shown and added by Fock. In addition to this exchange interaction, electrons also feel the change of each others Coulomb potential. As a result, the movement of one electron will result in a change of the Coulomb potential for all the other electrons in the system. This means that the motion of the electrons will be correlated. Such a correlation is not included in the Hartree-Fock (HF) approximation, but it is in higher levels of theory.\\
\indent The mean field concept of the HF approximation leads to a possible further level of abstraction: since electrons are mutually indistinguishable, and one is mainly interested in their collective behavior, it is possible to replace the electrons by their collective density distribution. As such, the electrons lose their identity as particle in this representation and a more probabilistic view emerges. Such an approach is followed in DFT,  in which the electron density is the central variable.\cite{fn:ConceptDFT} Consequently, the associated computational cost does no longer scale with the number of particles, but it scales with the grid size of the electron density instead. This allows for the investigation of systems which are at least one order of magnitude larger than what is usually possible within the HF approach.\\
\indent In $1964$, Hohenberg and Kohn were the first to formulate a DFT; their theory was based on two theorems:\cite{HohenbergKohn:PhysRev1964}
\begin{description}
  \item[Theorem 1] For a many-electron system in an external potential $V_{ext}$, this external potential, and consequently the total energy, is uniquely determined as a functional of the electron density $\rho(\mathbf{r})$.
  \item[Theorem 2] The exact ground state energy of a system in an external potential $V_{ext}$ is the variational global minimum of a universal energy functional $E[\rho(\mathbf{r})]$. The density that minimizes this functional is the exact ground state density.
\end{description}
In their work, Hohenberg and Kohn showed that, using the electron density as a basic variable, the total energy of an electron gas is a unique functional of this density, which includes both exchange \textbf{and} correlation contributions. However, the proof of the existence of such an energy functional is of little practical use, since its exact form is unknown. Later, this was partially resolved by Kohn and Sham. They proposed the existence of a non-interacting system described by the total energy functional
\begin{equation}
E_{KS}[\rho(\mathbf{r})]=T[\rho(\mathbf{r})] + V_{ext}[\mathbf{r}] + E^{H}[\rho(\mathbf{r})] + E^{xc}[\rho(\mathbf{r})],
\end{equation}
in which the first term represents the (non-interacting) kinetic energy contribution, the second term represents the contribution due to the external potential, the third term is due to the electron--electron Coulomb interaction (also known as the Hartree term) and the fourth term contains both the exchange and correlation contribution.\cite{fn:XCcorrgarbage} Exact analytical forms are known only for the first three terms. In the fourth term, it is the correlation contribution which lacks a general analytic form. As a result, this last term needs to be approximated. The resulting set of equations is known as the Hohenberg--Kohn--Sham (HKS) equations:
\begin{equation}
\Bigg\{\frac{-\hbar^2}{2m_e}\nabla^2 + V_{ext}(\mathbf{r}) + E^{H}[\rho(\mathbf{r})]
+ E^{xc}\big[\rho(\mathbf{r})\big] \Bigg\} \psi_i(\mathbf{r})=\varepsilon_i\psi_i(\mathbf{r}),
\end{equation}
with $ E^{H}[\rho(\mathbf{r})] = \frac{e^2}{4\pi\varepsilon_0}\int{\frac{ \rho(\mathbf{r}^{\prime})}{\|\mathbf{r}-\mathbf{r}^{\prime}\|}\mathrm{d}^3\mathbf{r}^{\prime}}$ and  $E^{xc}[\rho(\mathbf{r})]=\frac{\delta\{\rho(\mathbf{r})\varepsilon_{xc}[\rho(\mathbf{r})]\}}{\delta
\rho(\mathbf{r})}$. At this point, an expression is required for $E^{xc}$. In solid state calculations, one of the simplest approximations is the local density approximation (LDA) in which $E^{xc}$ is based on results obtained for the homogeneous electron gas. $E^{xc}$ consists of two parts, the exchange and the correlation part: $E^{xc}=E^{x}+E^{c}$. For the homogeneous electron gas the exact expression is known for the exchange part:\cite{DiracPAM:PCPRS1930, ParrRGYangW:1989DFT_AandM}
\begin{equation}
E^{x}[\rho(\mathrm{r})]=-\frac{3}{4}\sqrt[3]{\frac{3}{\pi}}\int{\rho(\mathbf{r})^{4/3}\mathrm{d}^3\mathbf{r}}.
\end{equation}
However, for the correlation part no analytic expression is known. Instead, a parameterized expression is used, originally presented by Perdew and Zunger, which fitted the results obtained in high quality Monte Carlo calculations on the homogeneous electron gas performed by Ceperley and Alder (CA).\cite{PerdewZunger_LDA:prb1981, CA:prl1980} Since this example of an exchange correlation functional is only a function of the density, it can be considered a zeroth-order approximation. A first-order approximation would also include the gradient of the density. This is the idea behind a generalized gradient approximation (GGA), for which the exchange correlation functional has the general form:
\begin{equation}
E^{xc}_{GGA}[\rho(\mathbf{r})]=\int{f(\rho(\mathbf{r}),\nabla(\rho(\mathbf{r})))\rho(\mathbf{r})\mathrm{d}^3\mathbf{r}}.
\end{equation}
Unlike LDA, there are many different kinds of GGA of which the formulation by Perdew, Burke and Ernzerhof (PBE) has become fashionable in the solid state community in recent years.\cite{PBE_1996prl}\\

\begin{table}[!t]
\caption{Comparison of the calculated lattice parameter $a$, the band gap and the cohesive energy E$_{\mathrm{coh}}$ of bulk germanium to their experimental values.}\label{table:TR_bulkGe}
\begin{ruledtabular}
\begin{tabular}{l|ccc}
& $a$ & band gap & E$_{\mathrm{coh}}$ \\
& (\AA) & (eV) & (eV) \\
\hline
exp.       & $5.6575^{a}$ & $0.89^{b}$ & $-3.85^{c}$  \\
LDA$^{d}$  & $5.6466$ & $0$ & $-4.616$  \\
LDA$^{e}$  & $5.66$   & $0$ & / \\
GGA$^{d}$  & $5.7785$ & $0$ & $-3.821$  \\
PBE        & $5.7618^{f}$ & $0^{g}$   & $-3.742^{f}$
\end{tabular}
\end{ruledtabular}
\begin{flushleft}
$^{a}$ Ref.~\onlinecite{webpage:Webelements}, $^{b}$ Ref.~\onlinecite{Pollmann:prb1993}, $^{c}$ Ref.~\onlinecite{Kittel_SolidStatePhysBook}, $^{d}$ Ref.~\onlinecite{VanpouckeDannyEP:2009ThesisNW}, $^{e}$ Ref.~\onlinecite{Pollmann:prb1993}, $^{f}$ Ref.~\onlinecite{LejaeghereK:2012arXiv1204.2733L}, $^{g}$ Ref.~\onlinecite{HybertsenMS:PhysRevB1986}.
\end{flushleft}
\end{table}

\indent DFT as presented here is strictly a ground state formalism, which means that the calculated energy of excited states is lower than in reality and as a result band gaps tend to be underestimated. However, the work done over the last decades has shown that DFT correctly describes the ground state properties of most systems using the functionals presented here. In case of bulk Ge, the underestimation of the band gap is an extreme case where it vanishes completely (\textit{cf.}~Table~\ref{table:TR_bulkGe}). It can be recovered using more advanced functionals, such as hybrid functionals, or techniques such as the GW approximation,\cite{Pollmann:prb1993} or simply by slightly compressing the crystal lattice in the calculation.\cite{VanpouckeDannyEP:2009ThesisNW} Despite this, other properties such as the lattice parameter and cohesive energy are calculated to be in good agreement with the experimental values, as is shown in Table~\ref{table:TR_bulkGe}. Furthermore, the experimental observation of the formation of dimer row reconstructions on the Ge(001) surface is supported by calculations. The small energy difference between the buckled dimers in the $b(2\times 1)$ and $c(4\times 2)$ reconstructions, of $0.07$ eV,\cite{Needels:prb88, VanpouckeDannyEP:2009ThesisNW} gives rise to fast oscillatory behavior of these surface dimers, providing them with a symmetric signature in STM images.\cite{vanHouseltA:JPhysCondensMatter2010} %\cite{Sato artikel in Arie seesaw: ref 7}
The calculated dimer lengths range from $2.38$ to $2.51$ \AA\ and buckling angles are in the range of $19$--$20^{\circ}$, which is also in good agreement with the experimentally reported values.\cite{Needels:prl87, Jenkins:jp96, Gay:prb99, VanpouckeDannyEP:2009ThesisNW} Finally, despite the vanishing band gap for bulk Ge, the existing gap states for the Ge(001) surface reconstructions are qualitatively present in calculated density of states (DOS).\cite{Pollmann:prl1995}
%%\subsubsection{Ge in DFT?}
%nee dat steken we erin hierboven in korte samenvatting?? of is dat er over??

\subsection{Scanning Tunneling Microscopy}
\subsubsection{Working principle and background}
\indent In surface science, one of the most important goals is the determination of the surface structure of a system. Everything else traces back to this fundamental knowledge. The ultimate dream here, from the experimental point of view, is to observe the surface at atomic resolution, showing the real-space positions of the atoms involved. The development of the STM by Gerd Binnig and Heinrich Rohrer in $1981$ at the IBM lab in Z\"{u}rich made this dream come true.\cite{Binnig:apl1982,Binnig:prl1982,Binnig:prl1983} It should not come as a surprise that this invention earned them the Nobel Prize in Physics in $1986$.\\
\begin{figure}[t]
  \begin{center}
  \includegraphics[width=8cm,keepaspectratio]{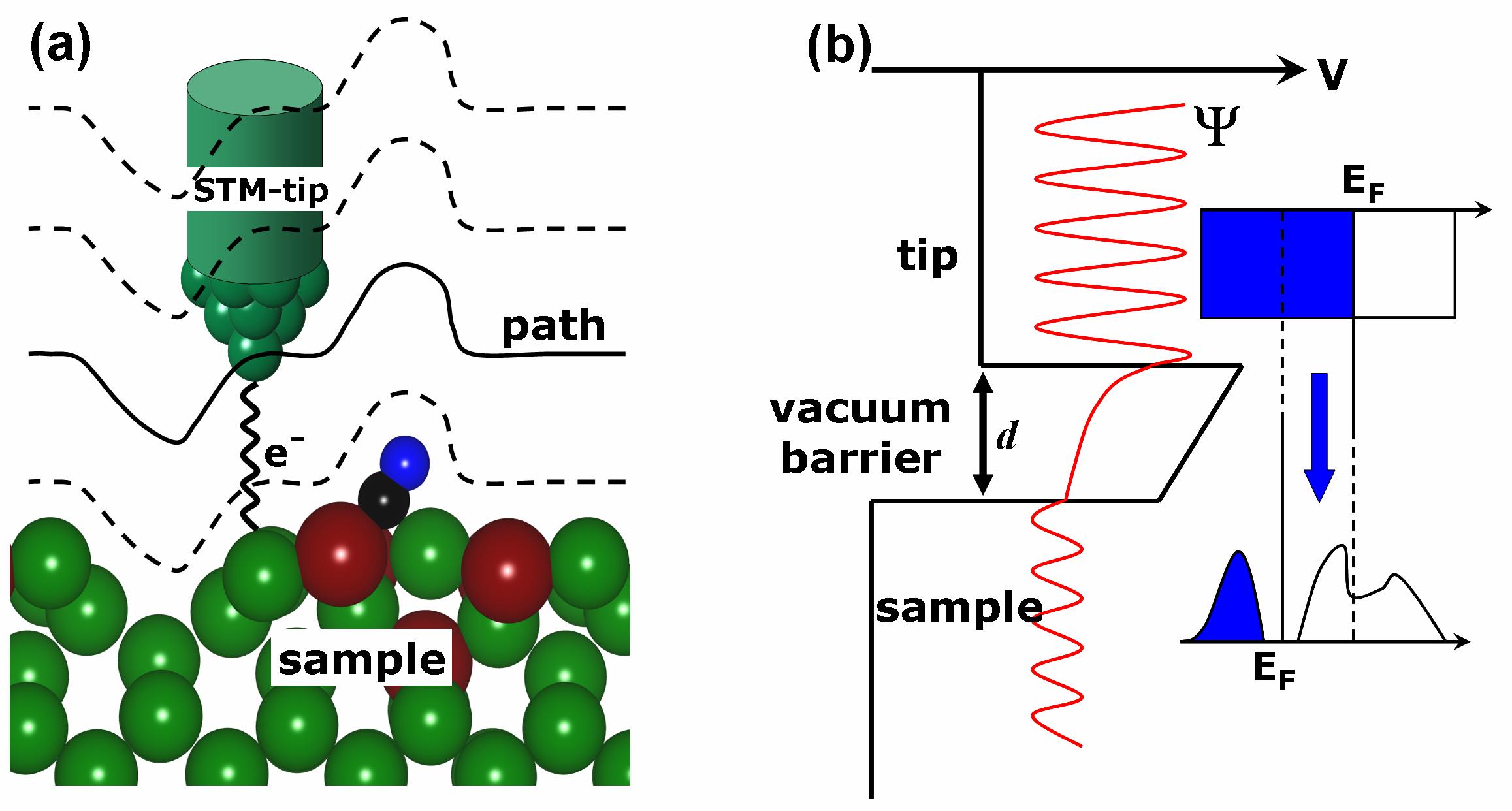}\\
  \caption{(a) Schematic representation of an STM experiment in constant current mode.
    An atomically sharp STM tip is brought close to the surface, such that electrons
    can tunnel between sample and tip. Keeping the current constant, the tip will
    follow a path parallel to the surface. (b) Quantum tunneling. An electron with
    incoming wave function $\Psi$ and energy close to the Fermi level of the tip,
    tunnels trough the vacuum barrier with thickness $d$ into an empty state
    in the sample.}\label{fig:TR_STMworking}
  \end{center}
\end{figure}
\indent An STM consists of a needle with an atomically sharp metallic tip. In general, tungsten or platinum-iridium is used, but also gold and recently carbon nanotubes have been used. This tip is brought very close to the surface, at a typical distance of $4$--$7$ \AA, and then scans it while measuring the tunnel current/resistance (\textit{cf.}~Fig.~\ref{fig:TR_STMworking}a). Because the tip does not physically touch the surface, a vacuum barrier between the tip and the surface exists, and electrons tunneling through this barrier give rise to a small but measurable current (\textit{cf.}~Fig.~\ref{fig:TR_STMworking}b). The probability of an electron with energy $E$ tunneling through a square barrier of height $V>E$ with a width $d$ is given, in the Wentzel--Kramers--Brillouin approximation by:
\begin{equation}
\mathrm{P}_{\mathrm{tr}}\cong e^{\alpha d},
\end{equation}
with $\alpha=\frac{-2}{\hbar}\sqrt{2m_e(V-E)}$. Due to this exponential relation, even a small difference in $d$ causes a large change in the tunneling probability, resulting in a high spatial resolution in the $z$-direction for an STM experiment.\\
\indent There are two modes of operation to scan a surface with an STM:
\begin{description}
  \item[Constant-height mode] In this mode, the STM tip is kept at a constant height while scanning over the surface. The change in tunnel current is measured, and resulting pictures give contour maps of this tunnel current. This allows for very fast scans, but it requires very flat surfaces as to prevent the tip from crashing into the surface.
  \item[Constant-current mode] In this mode, the current is kept constant using a feed-back-loop while scanning over the surface. In this case, not the tunnel current but the $z$-position is traced. The resulting images in this case are topographical maps of a constant current surface.(\textit{cf.}~Fig.~\ref{fig:TR_STMworking}a)
\end{description}
For the metal-induced NWs on Ge(001) discussed in this review, the latter mode of operation is generally used.
% If calculated and experimentally observed STM images are identical, one can very reasonably assume that the underlying geometry is the same. This argument is of the same class as: ``If it walks like a duck and talks like a duck, than it is most probably a duck'' \ldots\ though in some rare circumstances it turns out to be a platypus.\footnote{In \textbf{Chapter~\ref{ch_Nanowire}} we discover such a platypus.} This shows that the symbiosis between theory and experiment could make of STM an even stronger tool than it already is, by the introduction of chemical sensitivity. But also from the theoretical side it provides an additional powerful tool to study and compare theoretical models. One important success, realized already more than a decade ago, was the correct identification of the CO adsorption sites on Pt(111) from DFT calculations.\footnote{Further details on the history of this problem can be found in the introduction of \textbf{Chapter~\ref{ch_COmolecules}}.} \cite{Bocquet:ss96,Pedersen:cpl99}

\subsubsection[Tersoff-Hamann method]{Simulating STM: Tersoff-Hamann method}\label{ch_ss_THmethod}
The first theoretical description and simulation of STM by Tersoff and
Hamann is as nearly old as the experimental work itself.
\cite{Tersoff:prl83,Tersoff:prb85} They showed that the tunneling
current in an STM experiment is proportional to the local DOS (LDOS). A tunneling current is approximated by first-order perturbation theory in Bardeen's formalism: \cite{bardeen:prl1961}
\begin{equation}\label{eq_Current_perturb}
\mathrm{I}=\left(\frac{2\pi e}{\hbar}\right)\sum_{\mu,\nu} f(E_{\mu})\bigg[1-f(E_{\nu}+e\mathrm{V})\bigg]
|\mathrm{M}_{\mu\nu}|^{2}\delta(E_{\mu}-E_{\nu}),
\end{equation}
with $f(E)$ the Fermi function, V the applied bias,
$\mathrm{M}_{\mu\nu}$ the tunneling matrix between the tip states
$\psi_{\nu}$ (with energy $E_{\nu}$) and the surface states $\psi_{\mu}$
(with energy $E_{\mu}$). In the limit of small voltage and temperature,
Eq.~\eqref{eq_Current_perturb} reduces to:
\begin{equation}\label{eq_Current_limtempandvoltage}
\mathrm{I}=\left(\frac{2\pi e^{2}\mathrm{V}}{\hbar}\right)\sum_{\mu,\nu}
|\mathrm{M}_{\mu\nu}|^{2}\delta(E_{\mu}-E_{\mathrm{F}})\delta(E_{\nu}-E_{\mathrm{F}}),
\end{equation}
with the tunneling matrix element shown by Bardeen to be:
\cite{bardeen:prl1961}
\begin{equation}\label{eq_tunnelmatrix_bardeen}
\mathrm{M}_{\mu\nu}=\left(\frac{-\hbar^2}{2m}\right)\int{(\psi_{\mu}^{\ast}\nabla\psi_{\nu} -
\psi_{\nu}\nabla\psi_{\mu}^{\ast})\cdot\mathrm{d}\mathbf{S}},
\end{equation}
where the integral is over a surface in the barrier between the tip and the sample. Using a point-source as a simplified tip,
%\footnote{This approximation has some advantages from the theoretical point of view.Since the tip is practically ignored, there is no need for any information with regard to the tip geometry and calculated images will have a perfect resolution, making them ideal to distinguish surface features in experimental STM images from tip induced features.}
Eq.~\eqref{eq_Current_limtempandvoltage} reduces even further to:
\begin{equation}\label{eq_Current_pointsource}
\mathrm{I}\propto\sum_{\nu}|\psi_{\nu}(\mathbf{r}_0)|^2\delta(E_{\mu}-E_{\mathrm{F}}),
\end{equation}
which is the surface LDOS at $E_{\mathrm{F}}$. As a result, the experimentally measured surface of constant current is equivalent to a calculated surface of constant charge-density, from states close to $E_{\mathrm{F}}$. This most simple approximation is frequently used in theoretical work, since more advanced schemes require detailed structural information on the STM tip as well, which is generally not available. In addition, the theoretically infinitely sharp tip will allow for the calculated images to have a ``perfect'' resolution, making them a well-suited reference to distinguish surface features in experimental STM images from tip induced features and artifacts.\\
\begin{figure*}[t]
  \begin{center}
  \includegraphics[width=14cm,keepaspectratio]{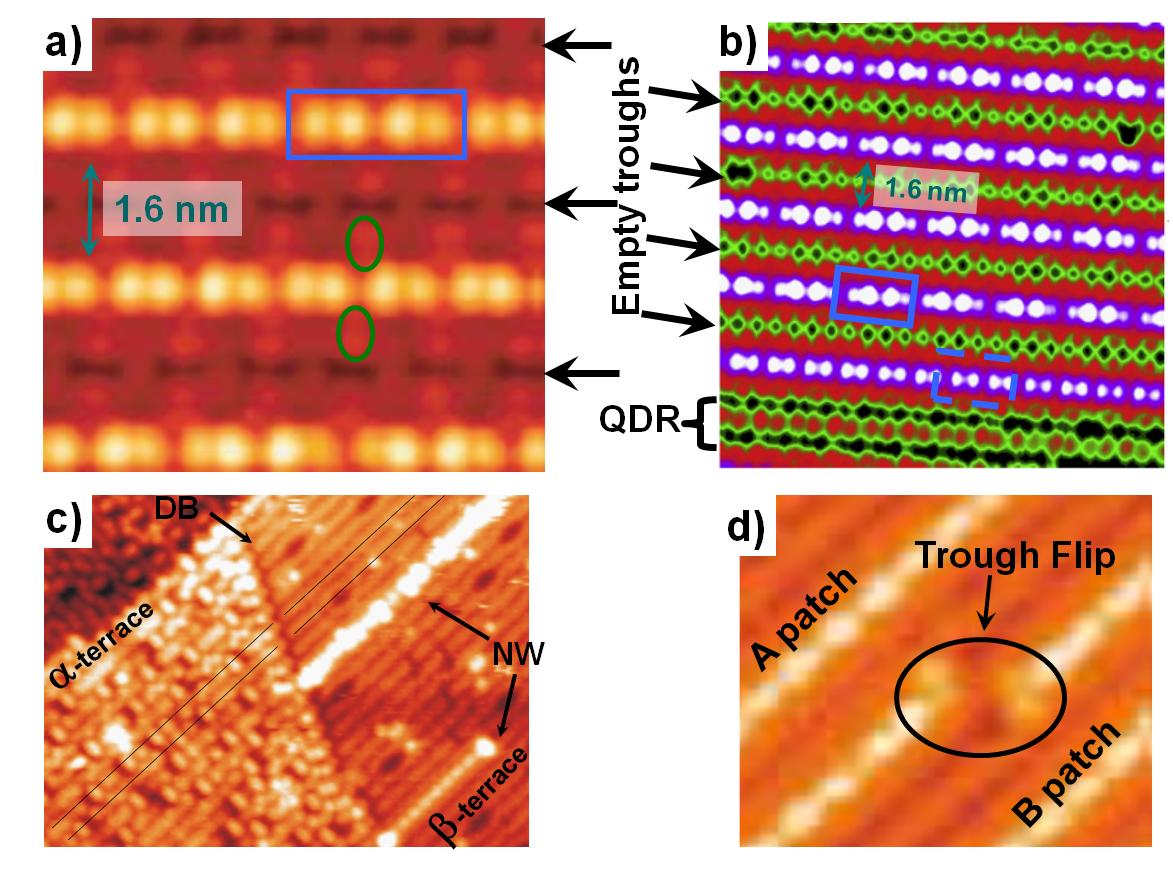}\\
  \caption{Typical experimental STM images of Pt NWs. (a) Empty state image taken at $77$ K, using a bias of $0.15$ V and a tunneling current of $0.437$ nA. Ellipses indicate the position of symmetric bulges observed near the NW. (b) Filled state images recorded at $4.7$ K, using a bias of $-1.5$ V and a tunneling current of $0.5$ nA. A QDR is visible at the bottom of the picture. (a and b) Empty troughs are indicated with arrows, while solid rectangles indicate pairs of NW dimers showing the ($4\times 1$) periodicity. The dashed rectangle indicates the ($2\times 1$) periodicity at the edge of a NW array. (c) A domain boundary between an $\alpha$-and $\beta$-terrace. The QDRs in both terraces are nicely aligned. (d) The boundary between two NW arrays (A- and B-patch) showing a trough flip. Note that each pair of NWs is spaced at least $1.6$ nm (or 1 empty trough) in the lateral direction. There is no interdigitation as observed for Au NWs on Ge(001). Figures are reproduced from Ref.~\onlinecite{VanpouckeDannyEP:2010bPhysRevB_NanowireLong}.}\label{fig:ExExpPtNW}
  \end{center}
\end{figure*}
\section{Pt nanowires on Ge(001)}
\subsection{Experimental background}\label{ss:PtNWsExp}
\indent In $2003$, G\"url\"u \textit{et al.} were the first to observe NWs formation on Ge(001) after deposition of roughly $0.25$ monolayer (ML) of Pt followed by ten minutes of annealing at $1050$ K.\cite{Gurlu:apl03} The observed one-atom thick wires could be up to a few hundred nanometers long and appeared both as solitary wires and in large arrays (\textit{cf.}~Fig.~\ref{fig:ExExpPtNW}). Their length is only limited by that of the underlying terrace, as is seen in Fig.~\ref{fig:ExExpPtNW}c: a Pt modified reconstructed Ge(001) terrace dubbed ``\emph{$\beta$-terrace}''.\cite{Gurlu:apl03, VanpouckeDannyEP:2009MaterResSocSympProc_Nanowire, VanpouckeDannyEP:2010aPhysRevB_BetaTerrace} This $\beta$-terrace has, at first glance, a simple dimer row structure. Upon closer inspection, a $c(4\times 2)$ symmetry is observed, with the dimer rows consisting of two distinctly different dimer-types. Therefore, these rows are named `\textit{quasi dimer rows}' (QDRs). On this $\beta$-terrace, the NWs fill the troughs between the QDRs.\\
\indent The NWs are defect- and kink-free, and the wire separation in the arrays is always exactly $1.6$ nm,\cite{fn:NWpt24nm} \textit{i.e.}\ every second trough of the underlying $\beta$-terrace. Conductivity measurements by G\"url\"u \textit{et al.} showed that the NW arrays are metallic in nature, which is consistent with the assumption that the NWs consist of Pt atoms.\\
\indent Upon closer examination of the NWs, all NWs appeared dimerised, leading to a basic ($2\times 1$) periodicity along the NWs. However, for NWs inside NW-arrays, a periodicity doubling was observed, leading to a ($4\times 1$) translational symmetry along the NWs (\textit{cf.}~Fig.~\ref{fig:ExExpPtNW}a and b). This ($4\times 1$) periodicity is observed to persist up to at least $77$ K, with a phase transition to ($2\times 1$) below room temperature.\cite{Houselt:ss08} In contrast, the ($4\times 1$) periodicity is never observed for solitary NWs and NWs at the edge of the NW-array. Also interesting is the fact that the buckling of adjacent NWs inside an array is always in-phase (\textit{cf.}~Fig.~\ref{fig:ExExpPtNW}b). This could be a hint that there is an interaction between the wires, either directly, via a two dimensional ($2$D) electronic interaction, or indirectly, via the substrate.\\
%\indent In a nutshell, metallic wires have been observed with a thickness of $0.4$ nm and lengths in the order of a micron: the dream of nearly any chip designer. These wires self-assemble, contrary to some earlier examples of monatomic wires that have to be built manually with an STM tip one atom at a time. They are also very robust: the observations by G\"url\"u \textit{et al.} were  done at room temperature (RT). Furthermore, when the NWs are exposed to the atmosphere, re-annealing them afterward in ultra high vacuum (UHV) showed the wires still to be intact.\\
\indent The experimental story, however, does not end here. In $2005$, \"Oncel \textit{et al.} presented the observation of quantum confinement between the Pt NWs.\cite{Oncel:prl05} These $1$D states, for which the NWs act as barriers, show an almost textbook behavior of a particle in a box. This is a somewhat surprising result in light of the earlier observation of the metallicity of the NW arrays by G\"url\"u \textit{et al.}, and the resulting tentative model. Considering the assumption that the wires consist of metal atoms, one would expect the wires to act as conductors rather than as barriers for these surface states. The same group also observed a very small band gap (BG) near the Fermi level, both at $77$ K and $300$ K, for an array of $1.6$ nm spaced NWs, contrary to the observation of G\"url\"u. The $dI/dV$ curve of the NW arrays given by G\"url\"u \textit{et al.}\cite{Gurlu:apl03} show an asymmetric dip toward zero near de Fermi level. The minimum value reached is of the order of $0.01$ nA/V, which could make both results consistent within the error.\\
\indent In $2006$, Sch\"afer \textit{et al.} \cite{Schafer:prb06} created Pt NWs on Ge(001) using a slightly lower anneal temperature of $600^{\circ}$C. Unlike \"Oncel \textit{et al.}, they found the presence of conduction states \textit{on} the wires. Also, later work by de Vries \textit{et al.} \cite{Vriesde:apl2008} seems to indicate that the NWs, with a ($2\times 1$) periodicity along the wire, are metallic. These observations and the coincidence of a small BG with the periodicity doubling leads van Houselt \textit{et al.} to interpret the ($4\times 1$) periodicity in terms of a Peierls instability.\cite{Houselt:ss08, ZandvlietHJW:JPhysCondMatter2009} The issue of the metallicity of the wires appears to be a difficult one, and the experimental observations point toward a quite complex electronic structure around the Fermi level.\\
\indent This is not the only complex behavior shown by the NWs. The most playful one to date is called the atomic pinball machine.\cite{Saedi:nl09} Saedi \textit{et al.} discovered that pairs of NW-dimers could be controlled using the current of the STM tip, flipping them back and forth like the flippers of a pinball machine.\\
\indent Of a more practical nature is the use of these nicely organized ultra-thin wires as a testbed for ($1$D) molecular electronics; for example by decorating them with selected molecules. In $2006$, \"Oncel \textit{et al.}\cite{Oncel:ss06} studied the diffusion and binding of CO on the NWs. They found that the CO coverage of the NWs is independent of pressure. Furthermore, only one adsorption site was observed, showing a protrusion in filled state STM images and a large depression in the empty state images. From this, \"Oncel \textit{et al.} concluded the adsorption site to be located on top of one of the NW dimer atoms, in line with the tentative model of Pt-dimers forming the NWs. The CO molecules also appeared to be highly mobile in this room temperature experiment, performing a $1$D random walk. Later experiments by Kockmann \textit{et al.}\cite{Kockmann:prb08} studied CO adsorption at $77$ K, discovering two more adsorption sites, located at the long and short NW dimer bridge positions (where short indicates the bridge position on a dimer, and long refers to the bridge site between two NW dimers). Contrary to the room temperature experiment by \"Oncel \textit{et al.}, the CO molecules remain immobile at $77$ K. Using statistical analysis of nearest neighbor spacings, a long range ($3$--$4$ nm) repulsive interaction between the CO molecules is revealed.\\
\indent \"Oncel and Kockmann chose to use CO for the specific reason that it has a high sticking probability and affinity for Pt while these properties are low for Ge. As such, CO serves as a means to characterize the NWs: preferential adsorption of CO on the NW would validate the tentative experimental model.\\
\indent Another example of molecular electronics is found in more recent work by Heimbuch \textit{et al.}\cite{HeimbuchR:PhysRevB2012, HeimbuchR:JPhysCondensMatter2013} They use an octanethiol molecule as a switch between their STM tip and a NW, and study the temperature dependence of the molecules' conductance. By varying the voltage across their molecular junction (tip--molecule--NW) they are able to controllably open and close the molecular switch.\\
\indent To this point, all experimental information available is linked (in some way) to STM-experiments. Recently Mochizuki \textit{et al.} provided experimental information on the atomic configuration and electronic structure using reflection high-energy positron diffraction (RHEPD) and angle-resolved photoemission spectroscopy (ARPES).\cite{MochizukiI:PhysRevB2012} This allowed them to test the, at the time available, theoretical and experimental models, so we will come back to their work in the following section.\\
\indent To our knowledge, only a single review of the experimental work on Pt NWs on Ge(001) exists, and it dates back to $2009$.\cite{SchaferJ:NewJPhys2009} The interested reader is referred to this  work by Sch\"{a}fer \textit{et al.} for more in depth details on the experimental setup of the different experiments and comparison to the Au nanowires on Ge(001), discussed in Sec.~\ref{sec:AuNWs}.
\begin{table*}[!tb]
\caption{Comparison of the different models for the Pt NWs on Ge(001). N$_{\mathrm{Pt}}$: the amount of Pt in the reconstruction. NW atoms: the atom type of the NW. STM: the simulation mode in the original papers is given; constant current (CC) or constant height (CH). E$_f^{\mathrm{LDA/GGA}}$: the LDA/GGA formation energies, respectively, given in meV/($1\times 1$) unit cell, using the Ge b($2\times 1$) as reference. As GGA functional the PW91 functional is used. r$_{dim}$: the dimer length of the NW dimer, as indicated in the original publications. The values presented are those taken from the references in the first column, unless indicated otherwise.}\label{table:TR_comparPtModels}
\begin{ruledtabular}
\begin{tabular}{l|rccccc}
 & N$_{\mathrm{Pt}}$ & NW atoms & STM  & E$_f^{\mathrm{LDA}}$ & E$_f^{\mathrm{GGA}}$ & r$_{dim}$ \\
& (ML) &  & & (meV) & (meV)& (\AA)\\
\hline
PDM\cite{Gurlu:apl03}
       & $0.50$ & Pt  & $-$ & $-$ & $-$  & $-$ \\
PNM\cite{SchwingenschloglU:EuroPhysJB2007, SchwingenschloglU:EPL2008}
       & $0.25$ & Pt  & $-^{a}$ & $-^{a}$ & $-^{a}$ & $3.50$\\
PNM$^{b}$ & $0.25$ & Pt  & $-$ & $+202$ & $+191$ & $2.70$\\
PINW1\cite{VanpouckeDannyEP:2008PhysRevB, VanpouckeDannyEP:2009MaterResSocSympProc_Nanowire, VanpouckeDannyEP:2010bPhysRevB_NanowireLong}
       & $0.75$ & Ge & CC & -257 & -370$^{c}$ & $2.72$\\
PINW2\cite{VanpouckeDannyEP:2009MaterResSocSympProc_Nanowire, VanpouckeDannyEP:2010bPhysRevB_NanowireLong}
       & $0.8125$ & Ge & CC & -364  & -458$^{c}$ & $2.65$ \\
TDC\cite{stekolnikov:prl08, StekolnikovAA:PhysRevB2008}
       & $0.25$  & Ge & CH & -56$^{d}$  & -92$^{d}$ & $2.55$\\
IPCM\cite{TsayShiowFon:SurfSci2012}
       & $0.25$  & Ge & CH & -159 & -158 & $2.60$ \\
\end{tabular}
\end{ruledtabular}
\begin{flushleft}
$^{a}$ This information is not provided by the authors.\\
$^{b}$ This work. Optimization of the the structure proposed in Ref.~\onlinecite{SchwingenschloglU:EuroPhysJB2007} showed that the Pt atoms in this model dimerize, and sink into the trough. As such, the formation energy shown here should be lower than that of the originally proposed structure.\\
$^{c}$ Taken from Ref.~\onlinecite{VanpouckeDannyEP:2009ThesisNW}.\\
$^{d}$ Taken from Ref.~\onlinecite{TsayShiowFon:SurfSci2012}.
\end{flushleft}
\end{table*}
\subsection{Theoretical models}
\begin{figure*}[!t]
  \begin{center}
  \includegraphics[width=15cm,keepaspectratio]{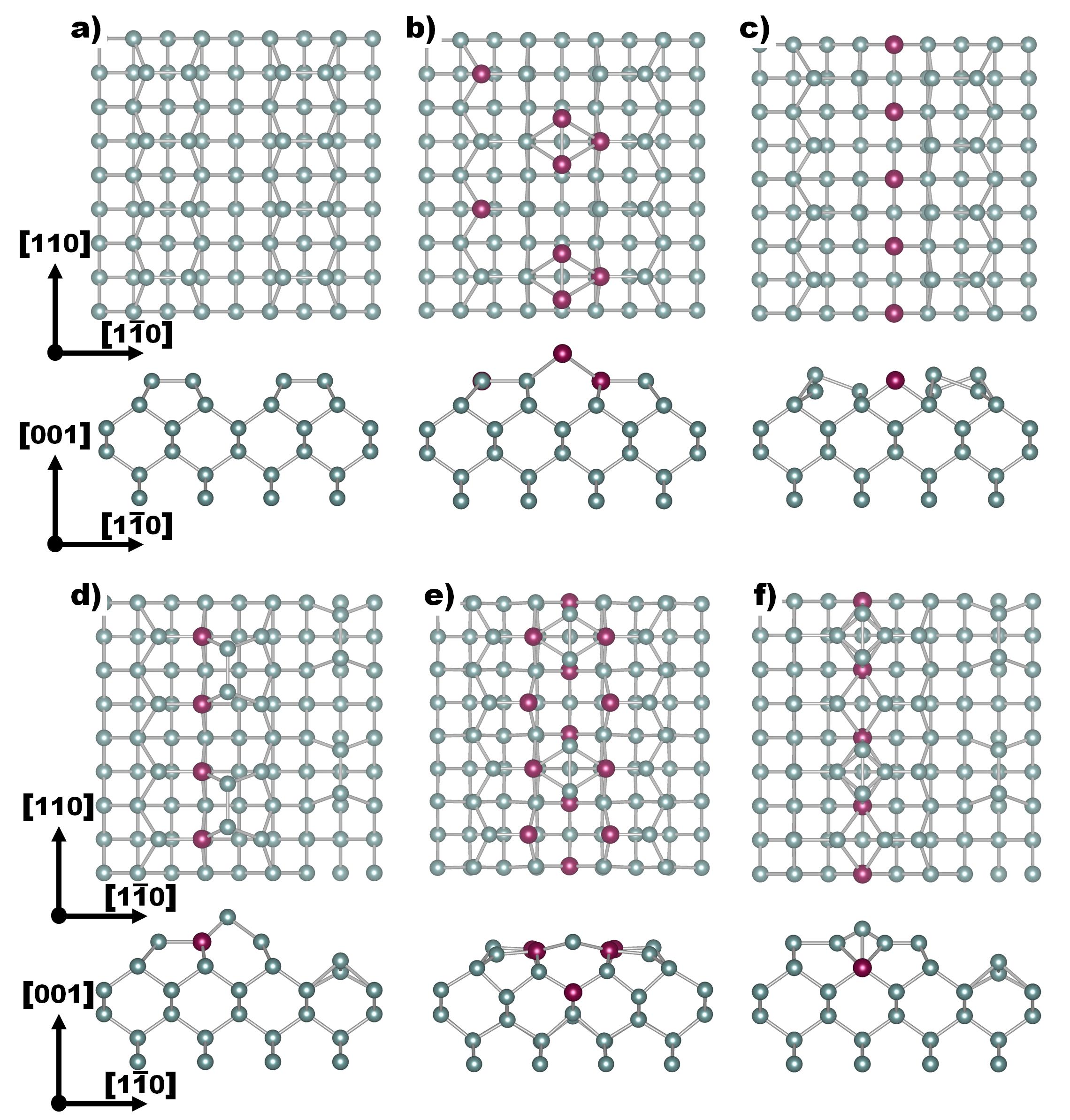}\\
  \caption{Ball-and-stick representations of the different models of Pt-induced NWs on Ge(001). Top: top view, bottom:side view. (a) Reconstructed Ge(001) surface as a reference, (b) PDM by G\"urlu et al.\cite{Gurlu:apl03}, (c) PNM by Schwingenschl\"{o}gl \textit{et al.}\cite{SchwingenschloglU:EuroPhysJB2007}, (d) TDC model by Stekolnikov \textit{et al.}\cite{stekolnikov:prl08, StekolnikovAA:PhysRevB2008}, (e) PINW model by Vanpoucke \textit{et al.}\cite{VanpouckeDannyEP:2008PhysRevB, VanpouckeDannyEP:2009MaterResSocSympProc_Nanowire, VanpouckeDannyEP:2010bPhysRevB_NanowireLong}, and (f) the IPCM by Tsay.\cite{TsayShiowFon:SurfSci2012}\\
  }\label{fig:TR_PtNWAllModels}
  \end{center}
\end{figure*}

\indent Despite all these beautiful experiments, the actual atomic structure of the system remains uncertain since STM is chemically insensitive. Although the gray- or red-scale images show the positions of atoms and dimers, they do not show which atomic species these atoms belong to. Atomic scale models on the other hand contain this information. Through comparison of simulated STM images of these models to experimental STM images, the actual atomic structure of the experimental system can be identified.\\
\indent Since the first experiments in $2003$, several models have been proposed (for reference, a model of the ($2\times 1$)-reconstructed Ge(001) surface with symmetric dimers is depicted in Fig.~\ref{fig:TR_PtNWAllModels}a):
\begin{description}
  \item[Pt-dimer model (PDM)] This is a tentative model proposed by G\"urlu et al.\cite{Gurlu:apl03} It consists of Pt-dimers in the troughs between de QDRs of a $\beta$-terrace and is shown in Fig.~\ref{fig:TR_PtNWAllModels}b.
  \item[Pt nanowire model (PNM)] A theoretical model which resembles the PDM, suggested by Schwingenschl\"{o}gl \textit{et al.}\cite{SchwingenschloglU:EuroPhysJB2007, SchwingenschloglU:EPL2008} In this model, shown in Fig.~\ref{fig:TR_PtNWAllModels}c, Pt-dimers with a dimer length of $3.5$ \AA\ are present between the dimer rows of a Ge(001) surface. Furthermore, every second Ge-dimer of every second dimer row is replaced by a single Ge atom.
  \item[Pt-induced nanowires (PINW)] Figure~\ref{fig:TR_PtNWAllModels}e shows the NW model presented by Vanpoucke \textit{et al.} in $2008$, based on their \textit{ab-initio} calculations.\cite{VanpouckeDannyEP:2008PhysRevB, VanpouckeDannyEP:2009MaterResSocSympProc_Nanowire, VanpouckeDannyEP:2010bPhysRevB_NanowireLong} In this model (PINW1), the NWs, consist of Ge-dimers located in the Pt-lined troughs between QDRs of Pt--Ge mixed-dimers. The atoms at the bottom of these specific troughs also consist of Pt atoms. Vanpoucke \textit{et al.} proposed this model for solitary Pt NWs, and the NWs at the edge of the NW arrays. A slightly modified model (PINW2) was proposed for the NWs in the arrays, which show a ($4\times 1$) periodicity along the NW. In this model, an additional Pt atom is located between every second pair of NW dimers, where a bond with these dimers gives rise to the tilted geometry of the dimers.\cite{VanpouckeDannyEP:2009MaterResSocSympProc_Nanowire, VanpouckeDannyEP:2010bPhysRevB_NanowireLong}
  \item[Tetramer-dimer-chain (TDC)] Based on their \textit{ab-initio} calculations, Stekolnikov \textit{et al.}\cite{stekolnikov:prl08, StekolnikovAA:PhysRevB2008} proposed a model showing a significant reconstruction of the surface top-layers. In this model, shown in Fig.~\ref{fig:TR_PtNWAllModels}d, every second dimer row is modified to consist of Pt--Ge mixed-dimers. The Ge-dimers of the dimer rows in between have broken up. They appear as if flipped over, on top of the Pt atoms of the Pt--Ge dimer row resulting in a NW consisting of Ge-dimers on top of the Pt atoms of the Pt--Ge dimer row. Second layer Ge atoms have also reconstructed to form Ge-dimers parallel to the NW, seen on the right side of Fig.~\ref{fig:TR_PtNWAllModels}d.
  \item[Imbedded Pt-chain model (IPCM)] The most recent theoretical model is proposed by Tsay, shown in Fig.~\ref{fig:TR_PtNWAllModels}f.\cite{TsayShiowFon:SurfSci2012} This model shows a surface reconstruction in which a Pt chain is half buried beneath two Ge-dimer rows. It can be derived from the TDC model by exchanging the Pt atoms and the second layer Ge atoms they are bound to. The Ge atoms making up the NW in the TDC model, retain their bonds with the Pt atom and as a result, the Pt NW is buried below two Ge-dimer rows separated a single Ge bond length. Between every pair of dimer rows covering a Pt chain, a trench dimer row exists, similar to the one found in the TDC model. The actual NW is formed by a Ge-dimer row, with the dimers oriented along the NW direction.
\end{description}
\indent Some structural details of the different models are summarized in Table~\ref{table:TR_comparPtModels}. It shows that both Pt- and Ge-dimers are proposed as the building blocks for the NWs, and the Pt content in the models varies from $0.25$ ML up to $0.875$ ML. The strengths and weaknesses of the models can be assessed by comparing the models to the experimental data.\\
\begin{table*}[!tb]
\caption{Calculated and estimated Pt concentration of the different terraces for Ge(001) surfaces with $0.25$--$0.6$ ML Pt deposited. Concentrations of Pt in the $\beta$-terrace and NW systems are those of the suggested models, Pt concentrations of the $\alpha$-terrace are estimated using Eq.~\eqref{eq_IL2_totdens} for samples with Pt deposition amounts of $0.25$--$0.30$ ML (indicated with subscript $1$) and $0.5$--$0.6$ ML (indicated with subscript $2$). For the latter case, the amount of Pt in the bulk is estimated under the assumption that the Pt concentration in the $\alpha$-terrace is that obtained for set $1$, as described in the text.\label{table:PtConcEstimate} }
\begin{ruledtabular}
\begin{tabular}{l|rrrrr}
 & $\rho_{\mathrm{Pt}}(\beta)$ & $\rho_{\mathrm{Pt}}(\mathrm{NW})$ & $\rho_{\mathrm{Pt}}(\alpha_1)$ & $\rho_{\mathrm{Pt}}(\alpha_2)$ & $M_{\mathrm{sub,2}}$ \\
  \hline
PDM            & $0.25$ & $0.50$  & $0.20\pm0.04$ & $2.46\pm0.50$ & $0.23\pm0.05$ \\
PNM, TDC, IPCM & $0.25$ & $0.25$  & $0.33\pm0.05$ & $3.25\pm0.50$ & $0.29\pm0.05$\\
PINW1          & $0.25$ & $0.75$  & $0.07\pm0.03$ & $1.68\pm0.50$ & $0.16\pm0.05$\\
PINW2          & $0.25$ & $0.8125$& $0.04\pm0.02$ & $1.48\pm0.50$ & $0.14\pm0.05$\\
\end{tabular}
\end{ruledtabular}
\end{table*}
\subsubsection{Experimental hints: Pt deposition and structure}
\begin{figure*}[!t]
  \includegraphics[width=12cm]{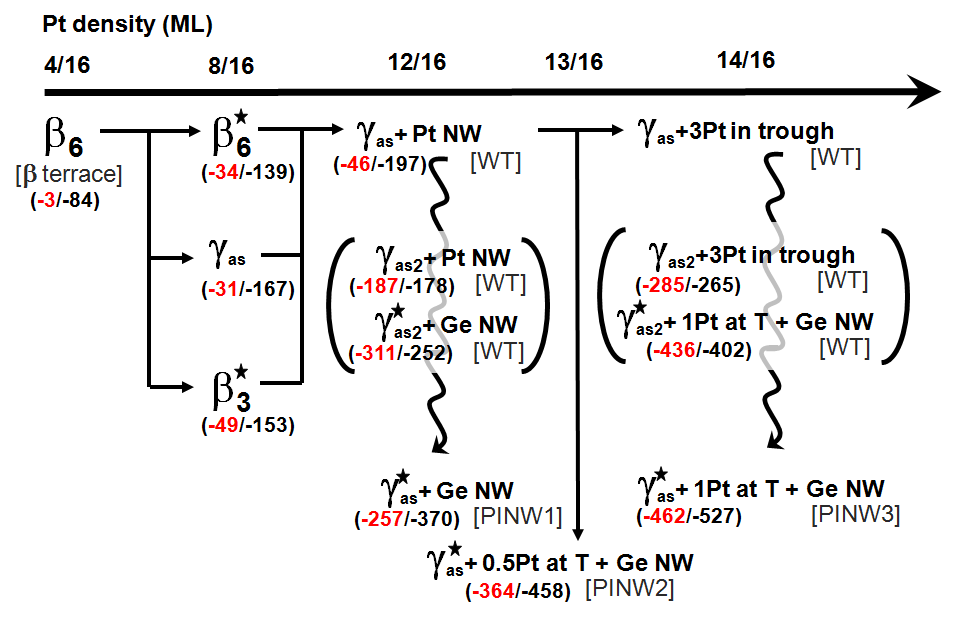}\\
  \caption{Formation paths proposed by Vanpoucke \textit{et al.} showing LDA/PW91 formation energies, in meV/($1\times 1$) surface unit cell, in red/black.\cite{VanpouckeDannyEP:2010bPhysRevB_NanowireLong, VanpouckeDannyEP:2009ThesisNW}}\label{Fig:FormationPathPtNW}
\end{figure*}

\begin{figure*}[!t]
  \includegraphics[width=12cm]{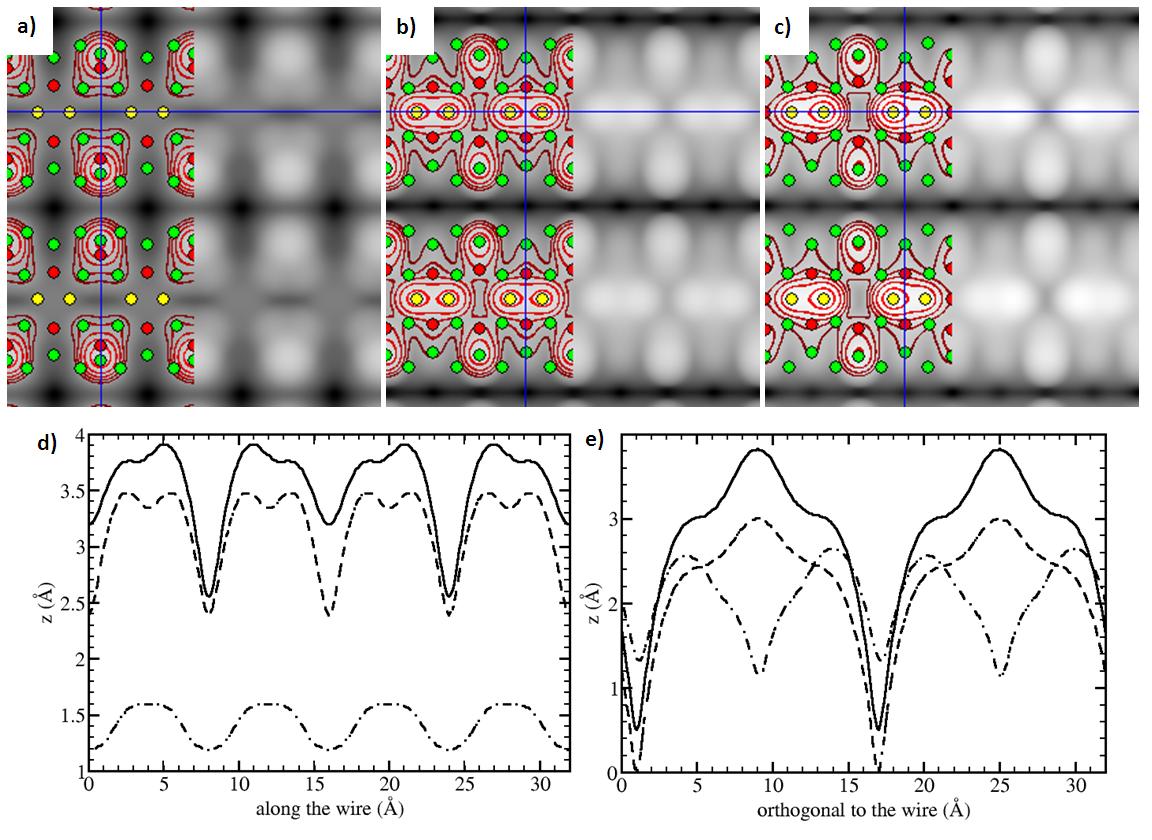}\\
  \caption{Calculated filled state STM images at a simulated bias of $-1.5$ V, of (a) a Pt NW in a Pt-lined trough of the Ge(001) reconstructed surface ($\gamma_{as}$+Pt NW in Fig.~\ref{Fig:FormationPathPtNW}), (b) PINW1, and (c) PINW2 showing a ($4\times 1$) periodicity along the NW. The green/red/yellow discs show the positions of the Ge/Pt/NW-atoms in these structures.\\ Cross-sections of the STM images (d) along and (e) orthogonal to the wire direction are given for the 3 systems: (a) dash-dotted line for the Pt NW, (b) dashed line for PINW1, and (c) solid line for PINW2. Figure taken from Ref.~\onlinecite{VanpouckeDannyEP:2009MaterResSocSympProc_Nanowire}, reprinted with permission.}\label{Fig:PINW_STM}
\end{figure*}

\indent A first piece of experimental data is the amount of Pt deposited. In most experimental work, this is stated as $0.25$ ML,\cite{Gurlu:apl03, Oncel:prl05, Fischer:prb07, Houselt:ss08} although some authors report larger deposition amounts of $0.5$ or even $1.2$ ML.\cite{MochizukiI:PhysRevB2012} Investigation of STM images presented shows the surface is only partially covered with Pt NWs. To obtain a surface fully covered with NWs, relatively large amounts of deposited Pt are needed.\cite{MochizukiI:PhysRevB2012} As such, one should consider $0.25$ ML as a minimum value rather than a target. However, several of the proposed models strictly adhere to a Pt content of $0.25$ ML (\textit{cf.}~Table~\ref{table:TR_comparPtModels}). It is interesting to observe that the PDM of G\"urlu contains $0.5$ ML of Pt.\cite{Gurlu:apl03} In the work of Vanpoucke \textit{et al.},\cite{VanpouckeDannyEP:2010bPhysRevB_NanowireLong} it is suggested that there exist gradients in the Pt content of these systems, allowing for much higher Pt content in the regions with the NW arrays (\textit{cf.}~Fig.~\ref{Fig:FormationPathPtNW}). At $0.25$ ML of Pt, this gives rise to the $\beta$-terrace on which the NWs form, similar as in the PDM. An increase to $0.75$ ML of Pt is required for the actual formation of NWs (PINW1). A further local increase to $0.8125$ ML (PINW2) gives rise to the formation of NW arrays, containing NWs with a ($4\times 1$) periodicity along the wire. The PINW1 do not show this ($4\times 1$) periodicity (\textit{cf.}~linescan images of simulated STM in Fig.~\ref{Fig:PINW_STM}). This is also the case for the experimentally observed solitary NWs and the NWs at the edge of the arrays, showing agreement with the suggestion of local gradients in the Pt concentration. From experimental observation, a surface hierarchy as function of Pt deposition is known to exist, which allows us to assume $\rho_{\mathrm{Pt}}(\alpha\mathrm{-terrace}) \leq \rho_{\mathrm{Pt}}(\beta\mathrm{-terrace}) \leq \rho_{\mathrm{Pt}}(\mathrm{NWs})$. Given the amount of deposited Pt, $M_{\mathrm{dep}}$, a relation with the surface phases can be given by:
\begin{equation}\label{eq_IL2_totdens}
M_{\mathrm{dep}}= \rho_{\mathrm{Pt}}(\alpha)X_{\alpha} +
\rho_{\mathrm{Pt}}(\beta)X_{\beta} +
\rho_{\mathrm{Pt}}(\mathrm{NW})X_{\mathrm{NW}} + M_{\mathrm{sub}},
\end{equation}
with $X_{I}$ the surface fraction covered by a terrace of type $I$ and $M_{\mathrm{sub}}$ the amount of Pt that is located deep in the subsurface, \textit{i.e.}\ below the third layer of \textit{all} the terraces. Since this equation has four unknown variables, four different sets ($M_{\mathrm{dep}}$, $X_{\alpha}$, $X_{\beta}$, $X_{\mathrm{NW}}$) suffice to find an exact solution for the local densities. To this date, no experimental work has been published presenting Pt densities of the different terraces. Based on two data sets containing estimates for surface coverage fraction of the different terraces, kindly provided by Prof. Zandvliet, it is possible to check the viability of the different models. Data set $1$ is
($M_{\mathrm{dep}} = 0.25$-$0.3$ ML, $X_{\alpha} = 30$\%, $X_{\beta} =
52.5$-$56$\%, $X_{\mathrm{NW}} = 14$-$17.5$\%) and data set $2$ is
($M_{\mathrm{dep}} = 0.5$-$0.6$ ML, $X_{\alpha} = 10$\%, $X_{\beta} =
58.5$\%, $X_{\mathrm{NW}} = 31.5$\%).\\
\indent Since the solubility of Pt in Ge is very small, the term $M_{\mathrm{sub}}$ can be assumed to be zero for low deposition amounts (data set $1$).\cite{Massalski:BAPD90}\\
\indent Using the first data set in Eq.~\eqref{eq_IL2_totdens}, the Pt content of the $\alpha$-terrace can be estimated for the different models. Table~\ref{table:PtConcEstimate} shows that the PDM and PINW models give rise to concentrations $<0.25$ ML (\textit{cf.}~$\rho_{\mathrm{Pt}}(\alpha_1)$), consistent with the assumption of gradients in the Pt concentration. The PNM, IPCM, and TDC models give rise to a Pt concentration of $0.33$ ML, which is consistent with the assumption in these models that the different terraces are different configurations with the same constant Pt concentration. The assumption of zero diffusion into the bulk cannot be used for the second data set, due to relatively high amount of deposited Pt, as is shown by $\rho_{\mathrm{Pt}}(\alpha_2)$ in Table~\ref{table:PtConcEstimate}. Instead, one can assume the Pt concentration in the $\alpha$-terrace to be the same as the one found for the first set, and estimate the amount of Pt diffusing into the bulk, which is shown in Table~\ref{table:PtConcEstimate} as $M_{\mathrm{sub}2}$. The values found for $M_{\mathrm{sub}2}$ are comparable to those found for the $\alpha$-terrace, and could thus be considered reasonable.\\
\indent When building a model for a surface NW, important information is extracted from the alignment of the NWs to the substrate. In this case, the substrate on which the NWs grow, the $\beta$-terrace, is a Pt-modified reconstructed Ge(001) surface.\cite{Gurlu:apl03} The resulting QDRs are not shifted with regard to the original Ge-dimer rows, so the topology of the c($4\times 2$) Ge(001) surface is retained, albeit with half of the Ge-dimers replaced by Pt--Ge heterodimers, giving rise to a checkerboard pattern.\cite{Gurlu:apl03, VanpouckeDannyEP:2010aPhysRevB_BetaTerrace} Among the proposed models, two categories can be distinguished: those in which the full dimer row topology of the original reconstructed Ge(001) surface is maintained (PDM, PINW), and those in which half of the dimer rows is significantly modified (PNM, TDC). The IPCM lies in between: since one dimer row is shifted half a dimer length, it can be considered missing at its original position. As such, the IPCM is the only model which does not retain the alignment of dimer rows underlying the NWs. This should show up in STM images of the $\beta$-terrace. Furthermore, the $\beta^{\ast}$-terrace suggested by Tsay, does not show the $c(4\times 2)$ symmetry observed for the experimental $\beta$-terrace, making it an unlikely candidate for this reconstruction. However, one cannot exclude the possibility that during the NW formation process, the $\beta$-terrace undergoes further reconstructions to form a $\beta^{\ast}$-terrace under the NW. On the other hand, experimental observation suggests that such a reconstruction may not be present. During an experimental study of the formation of Pt NWs, Fischer \textit{et al.}\cite{Fischer:prb07} noted that the formation of a widened trough preceding the formation of the NW. In this structure, the width of the trough between the QDRs is slightly increased, and only a single STM feature is present for every pair of dimers along the QDR. This reduction of the number of dimer features, is not found for the $\beta^{\ast}$-terrace of the IPCM. Alternately, in the work of Vanpoucke \textit{et al.}, this widened trough is suggested to be a surface reconstruction containing an extremely tilted Pt--Ge heterodimer sticking out of the surface.\\

\subsubsection{Theoretical STM images}
\indent Since STM images can be obtained experimentally and simulated for each of these theoretical models, it is an invaluable tool for straightforwardly comparing experiment and theory. In experiments, a constant-current mode is generally employed, resulting in height-maps of the surface. Although this is also easily possible in theoretical work, some authors (cf. Table~\ref{table:TR_comparPtModels}) opt to use a constant hight-mode approach, resulting in a density map. The theoretical models are diverse, but still two quite general and perhaps counterintuitive conclusions can be made based on the presented STM images. Firstly, the visible NW in the experimental STM image does not consist of Pt atoms. In the work of Vanpoucke \textit{et al.}\cite{VanpouckeDannyEP:2008PhysRevB, VanpouckeDannyEP:2009MaterResSocSympProc_Nanowire, VanpouckeDannyEP:2010bPhysRevB_NanowireLong} and Tsay \textit{et al.}\cite{TsayShiowFon:SurfSci2012}, it is shown that Pt atoms built into the surface remain invisible in STM (\textit{cf.}~Fig.~\ref{Fig:PINW_STM}a). Secondly, the theoretical models agree (except the PNM and PDM) on the NW seen in the STM images to be built out of Ge-dimers located on top of imbedded Pt atoms.\\
\indent The correct change of the NW image at varying bias is one of the more complicated properties of the NWs to reproduce. For a large negative bias and up to a small positive bias, the dimers clearly show a double peak indicating the atom positions, while for a large positive bias only a single dimer image is visible. From the theoretical studies, it is clear that the presence of Pt atoms right below the Ge NW dimer is essential for correctly reproducing this behavior.\\
\begin{figure*}[!t]
  \begin{center}
  \includegraphics[width=12cm,keepaspectratio]{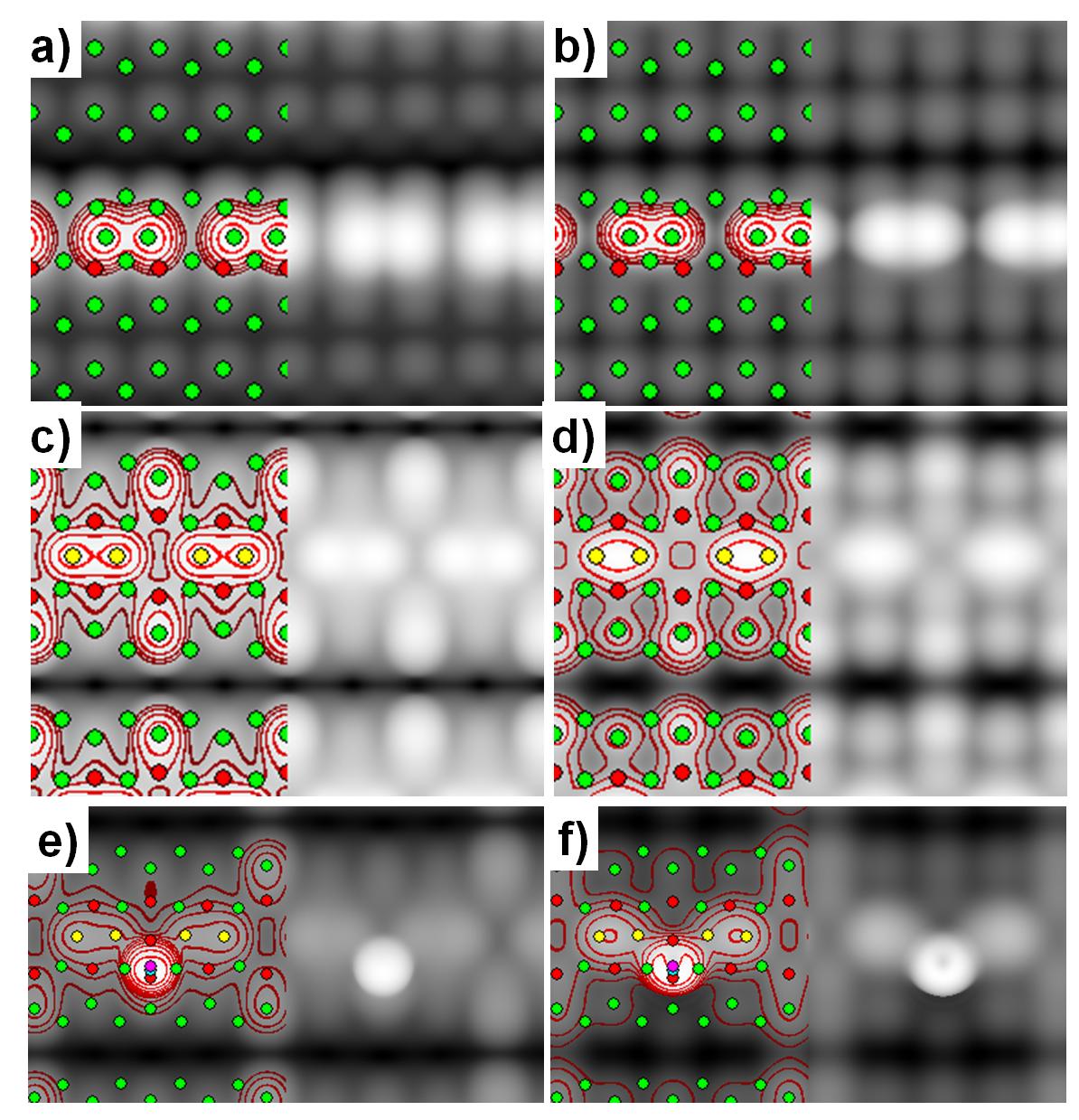}\\
  \caption{ (a,c) Filled and (b,d) empty state simulated STM images, at a simulated bias of $-1.50$ and $+1.50$ V, respectively, of (a,b) the relaxed Ge b($2\times 1$) T$_{1d}$ structure which resembles the TDC very closely, except for the reconstruction between the NWs, and of (c,d) the PINW1 model. (e) Filled and (f) empty state simulated STM images, at a simulated bias of $\pm1.80$ V, of a CO molecule adsorbed at a B2 site of the PINW2. Contours are added to guide the eye and red/green/yellow discs indicate the positions of the Pt/Ge surface/Ge NW atoms. Figures taken from Ref.~\onlinecite{VanpouckeDannyEP:2010bPhysRevB_NanowireLong} and \onlinecite{VanpouckeDannyEP:2010cPhysRevB_COonNW}.}\label{fig:STM_PtNW_CO}
  \end{center}
\end{figure*}

\subsubsection{CO adsorption}
\indent Already early on in the  study of Pt NWs, the adsorption of CO was investigated. The reason for this is twofold: firstly, as a means to support the experimentally suggested PDM; secondly, to study their use for molecular electronics applications. Under the assumption that the NWs consist of Pt atoms, CO molecules should only adsorb on the NW, and not onto the regions in between the NWs, which are supposed to consist of Ge atoms. The experimental observation of CO decorated NWs appeared to support the model perfectly, and appears to contradict models with Ge NWs. To this date, CO adsorption has been studied theoretically on free-standing Pt wires,\cite{SclauzeroG:PhysRevB2008} the Ge(001) c($4\times 2$) surface,\cite{HeHuiJing:SurfSci2012} the TDC,\cite{KrivosheevaAV:ComputMaterSci2010} and the PINW.\cite{VanpouckeDannyEP:2010cPhysRevB_COonNW} In their study of CO on Ge(001), He \textit{et al.}\cite{HeHuiJing:SurfSci2012} find a diffusion barrier of $0.607$ eV, and an adsorption energy of $0.628$ eV, making CO only weakly bound to the Ge surface, and suggesting that the diffusion process is a series of desorption-adsorption reactions. Combined with the observed $1$D random walk behavior in room temperature experiments, this would suggest it to be unlikely for the NWs to consist of Ge-dimers, as proposed in the TDC, PINW and IPCM.\cite{Oncel:ss06} For both the TDC and PINW, it is shown that CO adsorbs preferentially on the Pt atoms in the surface, and not onto the NW itself.\cite{KrivosheevaAV:ComputMaterSci2010, VanpouckeDannyEP:2010cPhysRevB_COonNW} The location of the Pt sites allows for easy and natural $1$D diffusion paths, in agreement with experimental observation. In the case of the TDC, the LDA adsorption energy is calculated to be $2.8$ eV, while for the PINW, LDA adsorption energies vary from $1.3$ to $2.9$ eV for varying Pt sites. For both NW models, the authors observe that the CO molecule shows a significant tilt. However, for the TDC the CO molecule tilts away from the NW, while for the PINW, the CO molecule tilts toward the NW.\cite{VanpouckeDannyEP:2010cPhysRevB_COonNW, KrivosheevaAV:ComputMaterSci2010} As a result, the calculated STM images show a CO image `on' the NW (\textit{cf.}~Fig.~\ref{fig:STM_PtNW_CO}). For the TDC, no calculated STM images are reported, however, since the tilt angle has the same value (TDC: 63.5$^{\circ}$, PINW: 57$^{\circ}$) though the opposite direction, it is plausible that the resulting STM image will show the CO image in between the NWs instead of on top of them, in contrast to experimental observation.\\
\begin{figure}[!t]
  \begin{center}
  \includegraphics[width=8cm,keepaspectratio]{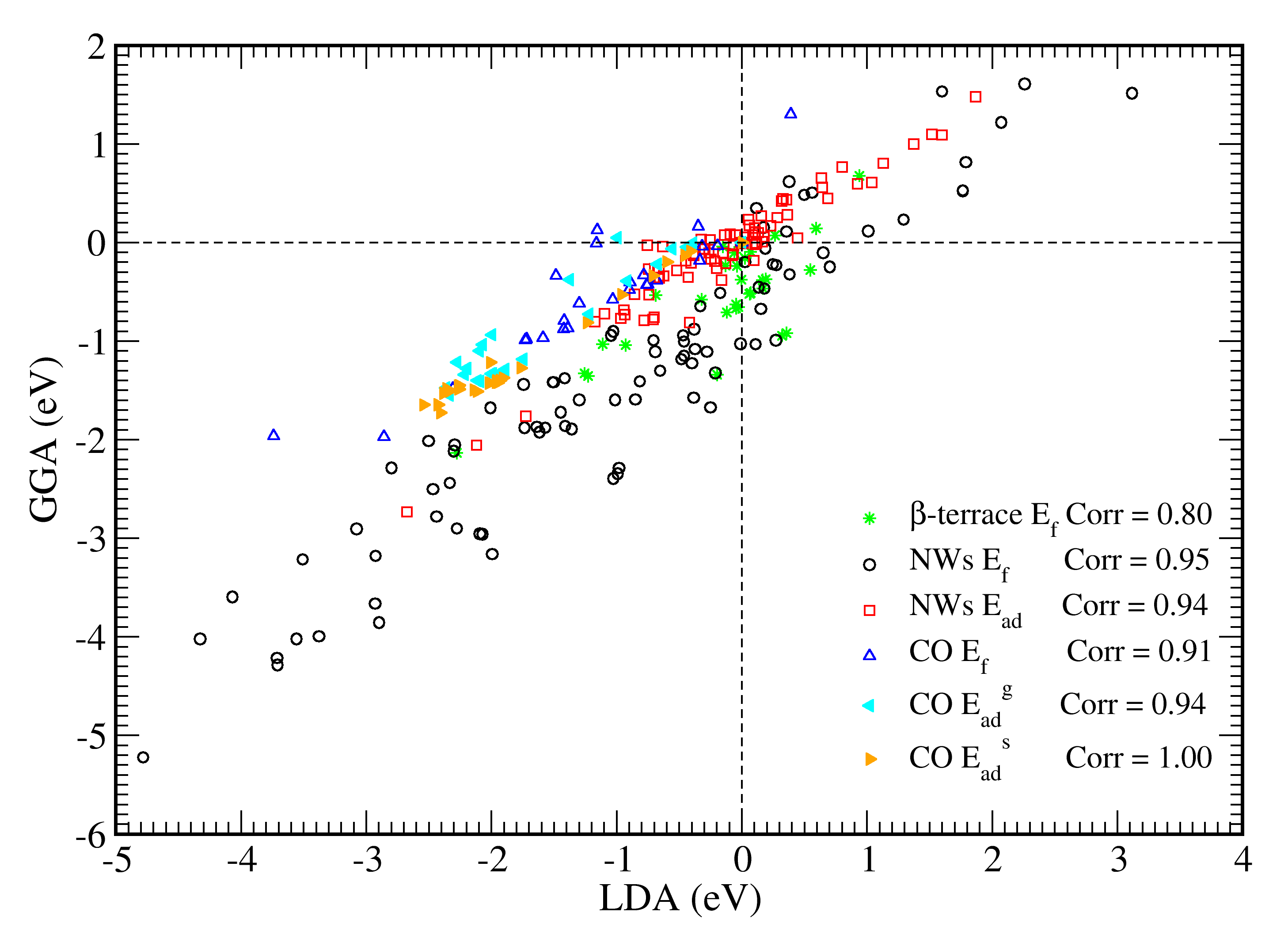}\\
  \caption{Correlation between LDA and GGA(=PW91) formation and adsorption energies. Figure taken from Ref.~\onlinecite{VanpouckeDannyEP:2009ThesisNW}}\label{fig:CorrCompLDAPBE}.
  \end{center}
\end{figure}

\subsubsection{Formation energy and electronic properties}
\indent In the aforementioned theoretical studies collectively, over $100$ structures of Pt modified Ge(001) have been investigated and relative formation energies have been presented. All studies show Pt to prefer subsurface positions, where more Pt--Ge bonds result in a more stable configuration, which is in agreement with the experimental observation of Pt moving into the substrate upon deposition. In the work of Vanpoucke\cite{VanpouckeDannyEP:2009ThesisNW} and Tsay\cite{TsayShiowFon:SurfSci2012}, it is shown that the same qualitative results are obtained for LDA and GGA functionals, shown for example by the correlation plot in Fig.~\ref{fig:CorrCompLDAPBE}.\\
\indent In Table~\ref{table:TR_comparPtModels}, the formation energies of the different NW models are shown. Although the IPCM, TDC, and PINW are all stable structures, there is a clear difference in stability of these structures. For systems limited to $0.25$ ML of Pt, the IPCM is the most stable configuration, with an energy comparable to that of Pt atoms buried in the third substrate layer under the Ge-dimer rows.\cite{TsayShiowFon:SurfSci2012, VanpouckeDannyEP:2010aPhysRevB_BetaTerrace} It is interesting to note that for the IPCM, the model for the $\beta$-terrace is more stable than the actual NW.\\
\indent For larger Pt concentrations, the PINW models are clearly the most stable configurations. The experimental PDM on the other hand was shown to be an unstable configuration, with the Pt-dimer breaking up and the Pt atoms burrowing into the substrate.\cite{VanpouckeDannyEP:2010bPhysRevB_NanowireLong}\\
\indent In a recent experimental study, Mochizuki \textit{et al.}\cite{MochizukiI:PhysRevB2012} performed reflection high-energy position diffraction (RHEPD) and angle-resolved photoemission spectroscopy (ARPES) measurements on the Pt NWs. They compared the obtained RHEPD rocking curves with simulated results for the PDM, TDC, PINW1 and PINW2. Of the models investigated, the PINW1 provided the best fit to their experimental curves. However, from their results it also followed that the Ge-dimers forming the NWs should be tilted at low temperature, with the tilt angle vanishing for temperatures above $110$ K. This tilt is not present in the work of Vanpoucke \textit{et al.}\cite{VanpouckeDannyEP:2010bPhysRevB_NanowireLong}, where it was noted that without additional atoms in the trough, the NW dimers optimize to flat symmetric dimmers.\\
\indent Experimentally, the transition to a ($4\times 1$) periodicity has been suggested to be a Peierls instability based on the observation of a significant reduction of the metallicity of the NWs.\cite{Houselt:ss08} However, recent ARPES experiments show no BG opens between the metallic bands at the Fermi energy, being an indication that no Peierls transition is  present.\cite{MochizukiI:PhysRevB2012, YajiK:PhysRevB2013} Moreover, Yaji \textit{et al.}\cite{YajiK:PhysRevB2013} suggest that the metallic band would be stable against a Peierls transition, in agreement with the findings of Vanpoucke \textit{et al.}\cite{VanpouckeDannyEP:2010bPhysRevB_NanowireLong}\\
%From their ARPES measurements, Mochizuki \textit{et al.}\cite{YajiK:PhysRevB2013} find two 1D metallic bands. It are the Pt atoms of the fourth layer in the PINW which contribute most to these bands and not the Ge atoms forming the NW. As such Mochizuki \textit{et al.} conclude these bands do not contribute to the structural phase transition of $2\times 1$ to $4\times 1$ periodicity of the NW. Even more, they conclude the bands to be stable against a Peierls instability if one would exist, in agreement with the findings of Vanpoucke \textit{et al.}\\
\subsection{Conclusion and outlook on Pt nanowires}
\indent Although multiple competing models for Pt NWs have been proposed, only three options remain after our previous evaluation: TDC, IPCM, and PINW. These three models do agree on some important points: (a) the experimentally observed NWs consist of Ge-dimers, (b) these Ge-dimers cover an imbedded 1D Pt reconstruction, (c) the formation of Pt--Ge bonds greatly stabilizes the structure.\\
\indent Each of the three models has its strengths and weaknesses. The TDC and IPCM present a reconstruction which naturally leads to a $\times 4$ periodicity orthogonal to the NW, which can easily be extended to a $\times 6$ periodicity giving rise to NW separations of $1.6$ and $2.4$ nm. The $\times 4$ periodicity along the NW is naturally linked to the Ge reconstruction in the trough between the NWs, but it will also produce this periodicity for solitary NWs. These models contain $0.25$ ML of Pt, which matches the $0.25$ ML deposition in the original experiments, although the latter statement does clearly not imply either a full coverage of the sample with NWs, nor a homogeneous distribution of the Pt. In fact, the experimental images clearly show partial coverage of the substrate with NWs, in addition to the $\beta$- and $\alpha$-terraces.\cite{Gurlu:apl03} For the PINW, the origin of the $\times 4$ periodicity orthogonal to the NW is not present as explicitly. In this model, the instability of imbedded Pt homodimers in the  surface prevents the formation of NWs in adjacent troughs.\cite{VanpouckeDannyEP:2010aPhysRevB_BetaTerrace} As a result, only two types of NW spacings are expected: $1.6$ nm inside single NW arrays, and $2.4$ nm where two NW arrays mutually touch on the sides. In the PINW model, also trough flips---as are shown in Fig.~\ref{fig:ExExpPtNW}---can easily be explained, and would induce only limited stress into the surface reconstruction, due to the Pt--Ge Ge--Pt dimer interaction in the QDR at the trough flip.\cite{VanpouckeDannyEP:2010aPhysRevB_BetaTerrace}\\
\indent The TDC, PINW, and IPCM give rise to STM images which resemble the experimentally observed NWs, although some of the more subtle features differ. For both the PINW and IPCM, a widened trough model is proposed, although in the IPCM the experimentally observed halving of features in the QDR (one feature for every pair of dimers) is not observed, in contrast to the PINW. The study of CO adsorption on the TDC and PINW shows similar behavior for the CO molecules: a tilt of about $30^{\circ}$ away from the surface normal. However, for the PINW the tilt is toward the NW, whereas it is away from the NW for the TDC, allowing for better agreement of the former to the experiment. The IPCM, on the other hand, would be a bit problematic for CO adsorption, since the CO molecule cannot access the Pt buried in the substrate. As a result, the CO molecules would need to adsorb on the Ge NW dimers. This makes the experimentally observed 1D random walk along the NW very unlikely, since it is known from calculations on pure Ge(001) that CO diffusion would entail desorption-adsorption events, making inter-NW hopping much more likely than for the TDC and PINW.\\
\indent Of the models presented in this review, the PINW appears to explain most experimental STM observations, and to be in agreement with other experimental findings. Furthermore, recent ARPES and RHEPD rocking curves seem to match best for the PINW model, and support their prediction that the ($4\times 1$) periodicity is not due to the presence of a Peierls instability.\\
\indent Further improvement and testing of this model is needed, since some issues remain unresolved including: (a) the ($4\times 1$) periodicity of the NW (electronic or structurally induced; in the latter case, which other atoms may be involved?), (b) the exact Pt concentration in the different experimental reconstructions ($\alpha$-, $\beta$-terrace, NWs both solitary and in arrays, are there other phases inside the arrays which may indicate even higher Pt concentrations?) (c) the charge transfer between the Pt and Ge atoms in the NW, and (d) adsorption sites on the NW for other molecules, and the stability of the NW under adsorption.\\
\indent Point (b) is of special interest to theoreticians, since an answer to this reduces the phase space of the model search dramatically. To this date, most theoreticians have limited themselves to $0.25$ ML, while our simple estimates show that larger Pt concentrations are also consistent with observed surface fractions for the various terraces.\\

\begin{figure*}[!t]
  \begin{center}
  \includegraphics[width=16cm,keepaspectratio]{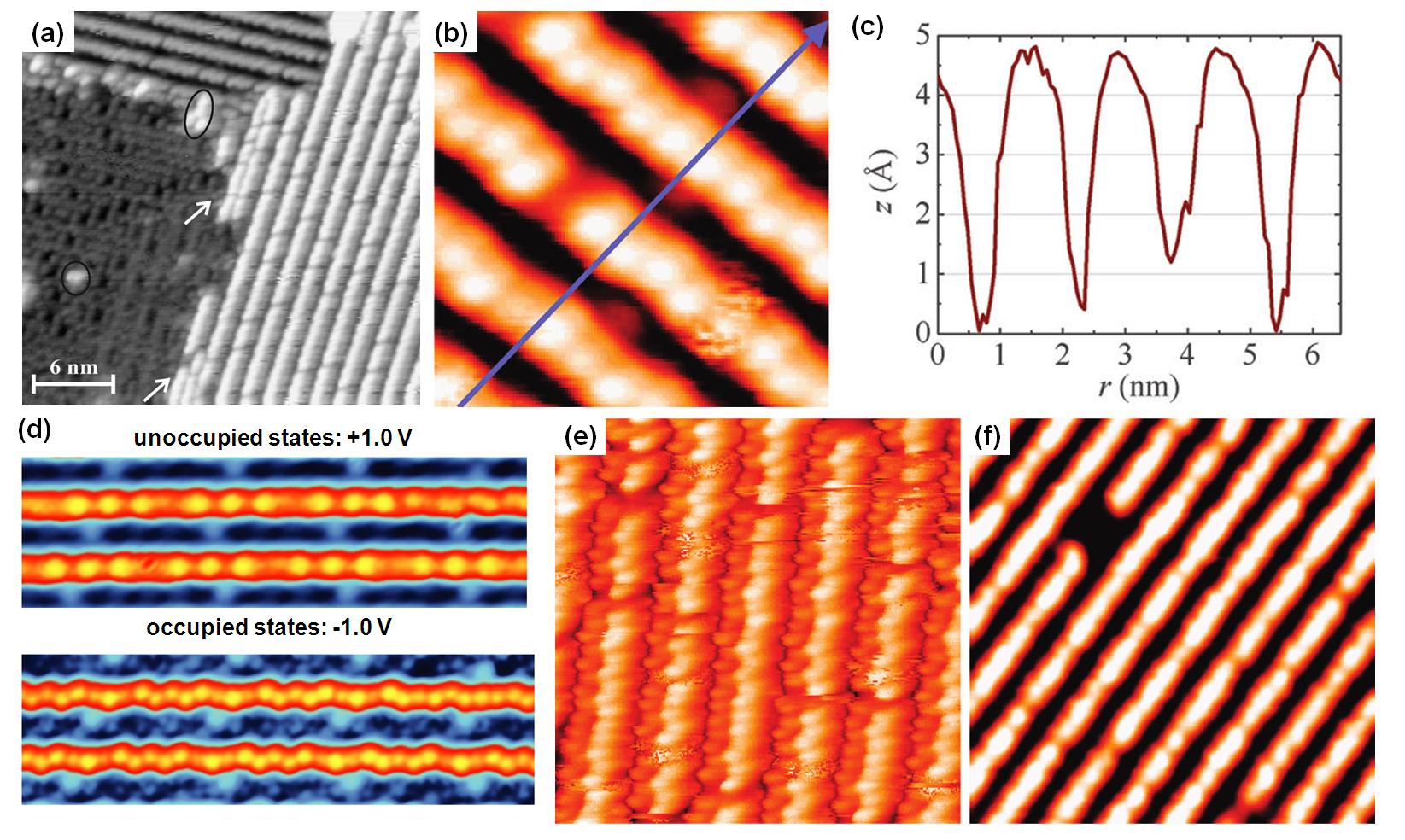}\\
  % fig 4 from Kockmann 2009...and one of the originals of Wang and SChaffer...to coompare to fig 4 Kockman??
  \caption{Typical STM images of Au NW. (a) Room temperature STM image at a sample bias of $-1.0$ V after deposition of an additional $0.05$ ML of Au on a sample with $0.5$ ML of Au deposited at $675$ K. The arrows and the ellipse indicate positions where the NWs show a spacing of only $8$ \AA\ instead of the usual $16$ \AA. (b) Occupied state STM image of Au NWs at room temperature for a sample bias of $-1.0$ V. (c) Linescan along the blue line indicated in (b) showing very deep troughs are present between the Au NWs. (d) Low temperature occupied and unoccupied STM images of Au NWs on Ge(001) showing a more detailed structure with a long range periodicity. (e) and (f) STM images of Au NWs prepared by deposition at $775$ K, showing how experimental STM-tip parameter influence the obtained image. e) resembles the Au NWs shown in (a) and is obtained with a sample bias of $-0.9$ V and a tunneling current of $3.0$ nA, while (f) resembles (b) and is obtained using a tunneling current of $0.4$ nA and a sample bias of $0.7$ V.\\
  Figures taken from Ref.~\onlinecite{WangJ:SurfSci2005} (a), Ref.~\onlinecite{KockmannD:JPhysChemC2009} (b,c,e,f) and Ref.~\onlinecite{BlumensteinC:JPhysCondensMatter2013} (d).}\label{fig:AuNW_STM}
  \end{center}
\end{figure*}
\section{Au nanowires on Ge(001)}\label{sec:AuNWs}
\subsection{Experimental background}\label{ss:AuNWsExp}
\indent In $2004$, one year after the first observation of Pt-induced NWs on Ge(001), Wang \textit{et al.}\cite{WangJ:PhysRevB2004} observed the formation of atomic scale NWs on a Ge(001) substrate after Au had been deposited. Similar as for the Pt NWs, these Au NWs are obtained after the deposition of $0.5$--$1.5$ ML of Au and show huge aspect ratios. Lower deposition amounts of only $0.1$ ML Au lead to the formation of vacancy riddled Ge(001) terraces, comparable to the $\alpha$-terrace during Pt NW formation. On these terraces, no NWs are found, indicating all the Au moved into the subsurface.\cite{Gurlu:apl03, WangJ:PhysRevB2004, WangJ:SurfSci2005, GallagherMC:PhysRevB2011, SafaeiA:PhysRevB2013} The observed dimer vacancies (DV) did not show any long range order, as is seen for Ni- and Ag-induced vacancy lines on Si(001).\cite{ZandvlietHJW:PhysRevLett1995, UkraintsevVA:SurfSci1996, ChangCS:SurfSci1996, ZandvlietHJW:SurfSci1997} However, they do form short-range DV complexes: (1+2+1) DVs along the dimer rows of the reconstructed Ge surface. The (1+2+1) DVs consist of a double DV in the middle, with a single DV on each side separated by a single dimer from the central double DV. For the Pt modified system, it was shown by means of \textit{ab-initio} calculations that such DV complexes do not necessarily indicate missing dimers but can also be produced by the presence of subsurface metal atoms (Pt or Au in these cases) modifying the electronic structure of the surface dimers in such a way as to make them `\textit{invisible}' for STM.\cite{VanpouckeDannyEP:2010aPhysRevB_BetaTerrace}\\
\indent For deposition amounts starting at $0.4$ ML, Wang \textit{et al.}\cite{WangJ:SurfSci2005} observed islands containing white and gray chains or NWs, shown in Fig.~\ref{fig:AuNW_STM}a. For increasing deposition quantities the islands grow larger, until the entire substrate is covered. Later growth studies by Melnik and Gallagher,\cite{GallagherMC:PhysRevB2011, MelnikS:SurfSci2012} and Safaei \textit{et al.}\cite{SafaeiA:PhysRevB2013} show the first small patches of Au NW already appear at a deposition amount of $0.1$ ML, although the fraction of NW covered surface is then still to small to be picked up in low-energy electron diffraction (LEED) experiments. Full surface coverage is reached at a deposition of $0.75$ ML. For higher coverage, the excess Au atoms form 3D Au islands on top of the NWs. Furthermore, Safaei \textit{et al.}\cite{SafaeiA:PhysRevB2013} showed that the formation of Au NW domains is driven by a dewetting-wetting transition at $665$ K. At temperatures between $320$ and $585$ K, the Au atoms diffuse away from the NW domains, while for temperatures above $665$ K Au atoms of the NWs start to form 3D islands, and for temperatures above $890$ K all Au diffuses into the bulk.\cite{SafaeiA:PhysRevB2013, GallagherMC:PhysRevB2011, WangJ:PhysRevB2004, WangJ:SurfSci2005} This is quite different from the Pt NWs, where the formation only occurs at such high temperatures.\cite{Gurlu:apl03}\\
\indent The Au chains are found to follow the topology of the original Ge substrate. Furthermore, Gallagher \textit{et al.}\cite{GallagherMC:PhysRevB2011} observed that at low deposition amounts, the NWs are atomically flat, even though they may cover multiple Ge terraces. This indicates that the growth of the NWs is accompanied by a significant mass transport of both Au and Ge. In the original work of Wang \textit{et al.}\cite{WangJ:PhysRevB2004}, the NWs are assumed to consist of Au-modified dimer rows, where the white chains consist of Au-dimers, while the gray chains consist of Au--Ge heterodimers. As such, the Au NWs may, just like the Pt NWs, be considered as examples of \textit{template-driven self-organization}.\cite{SchaferJ:NewJPhys2009} The alternating nature of the chains, and the zigzag appearance of the chains tops leads to a $(4\times 2)$ reconstruction supported by LEED experiments.\cite{WangJ:PhysRevB2004} This $(4\times 2)$ LEED pattern is also supported by the STM observation of alternating dimer rows with a white and gray chain character. However, at the borders of such arrays after additional Au deposition, or at higher deposition amounts, \textit{interdigitation} of the white chains is often observed (indicated in Fig.~\ref{fig:AuNW_STM}a by the white arrows).\cite{WangJ:PhysRevB2004, SchaferJ:NewJPhys2009} In such an interdigitated region, all dimer rows are replaced by white chains, making this system quite different from the Pt NWs, where such behavior is never observed.\\
\indent In $2008$, the groups of Zandvliet and Claessen simultaneously presented their observation of Au NWs on Ge(001) showing, at first glance, quite different results than those presented earlier by Wang \textit{et al.}\cite{Schafer:prl2008, vanHouseltA:PhysRevB2008}, this can be seen in Fig.~\ref{fig:AuNW_STM}b. Sch\"{a}fer \textit{et al.}\cite{Schafer:prl2008} only deposited $0.5$ ML of Au on a sample kept at $773$ K, and observed the formation of Au chains with a $c(8\times 2)$ periodicity covering the substrate. They suggest that the low temperature of the earlier experiments leads to the need for larger deposition amounts to obtain the same NW coverage. Also in contrast to the work of Wang \textit{et al.}\cite{WangJ:PhysRevB2004, WangJ:SurfSci2005} and of van Houselt \textit{et al.}\cite{vanHouseltA:PhysRevB2008}, Sch\"{a}fer \textit{et al.}\cite{Schafer:prl2008, SchaferJ:NewJPhys2009} deduce the chain width to be only a single (Au) atom and not a dimer. Furthermore, they claim the observation of a 1D metallic state along the Au NW (\textit{cf.} Sec.~\ref{sss:AuNW_controversies}). This conduction path is suggested to be decoupled from the substrate giving rise to a truly 1D electron liquid. The delocalized nature of the electrons on top of the NW results in STM images containing very little structural information (cf.~Fig.~\ref{fig:AuNW_STM}b), making modeling work harder than was the case for the Pt NWs. At the same time, this makes it a perfect toy  system for more abstract theoretical models dealing with exotic low dimensional physics.\\
\indent In the work of van Houselt \textit{et al.}\cite{vanHouseltA:PhysRevB2008}, $0.2$--$0.3$ ML of Au was deposited at room temperature, and annealed afterward at a temperature of about $650$ K. The resulting Au NWs are separated $1.6$ nm, just as observed by Wang \textit{et al.} and Sch\"{a}fer \textit{et al.} Their height, however, is measured to be at least $6$ \AA\ (four to five times the previously observed height), or $4$ Ge step heights (\textit{cf.}~Fig.~\ref{fig:AuNW_STM}c). As a result, Van Houselt \textit{et al.} propose what they call a giant missing row (GMR) reconstruction. In this reconstruction, shown in Fig.~\ref{fig:AuNWmodels_GMR}, the NWs are actually (111) microfacets of Ge, which are covered with Au trimers. In a reconstruction as shown in Fig.~\ref{fig:AuNWmodels_GMR}, about $1.5$ ML of Au would be required for full coverage of the substrate, in agreement with the results of Wang \textit{et al.}\cite{WangJ:PhysRevB2004} and in contrast to the findings of Sch\"{a}fer \textit{et al.}\cite{Schafer:prl2008}, but also at odds with later findings where full surface coverage at $0.5$ ML or $0.75$ ML deposition was obtained.\cite{MockingTF:SurfSci2010, GallagherMC:PhysRevB2011, MelnikS:SurfSci2012} On top of this ridge, a dimer row of anti-ferromagnetic buckled Ge-dimers is found, giving rise to the zigzag features in the STM images. The extreme height and sharpness of the NW feature requires an extremely sharp STM tip, without which the height of the NWs may well appear much lower---in agreement with the earlier work.\\
\indent At this point, it appears as though several different types of Au NW can be produced using different experimental parameters, which is not an unusual phenomenon.\cite{RossignolS:JMaterChem1999, VanpouckeDannyEP:2011PhysRevB_LCO} Kockmann \textit{et al.}\cite{KockmannD:JPhysChemC2009} and later Niikura \textit{et al.}\cite{NiikuraR:PhysRevB2011}, however, showed this not to be the case for Au-induced NWs on Ge(001). Kockmann \textit{et al.}\cite{KockmannD:JPhysChemC2009} showed that, since STM images are a convolution of the surface and the STM tip's electronic structure, the shape of the  latter will lead to a variety of STM images (\textit{cf.}~Fig.~\ref{fig:AuNW_STM}). By using a set of different STM tips to image their Au NWs, they were able to reproduce images which resemble either those presented in the work of Wang \textit{et al.}\cite{WangJ:PhysRevB2004} or in the work of Sch\"{a}fer \textit{et al.}\cite{Schafer:prl2008}, as is shown in Fig.~\ref{fig:AuNW_STM}e and f. They also show the NWs to have a width comparable to that of a dimer on the Ge(001) surface.\\
\subsection{Theoretical models}
\subsubsection{Experimental controversies}\label{sss:AuNW_controversies}
\begin{figure*}[!t]
  \begin{center}
  \includegraphics[width=16cm,keepaspectratio]{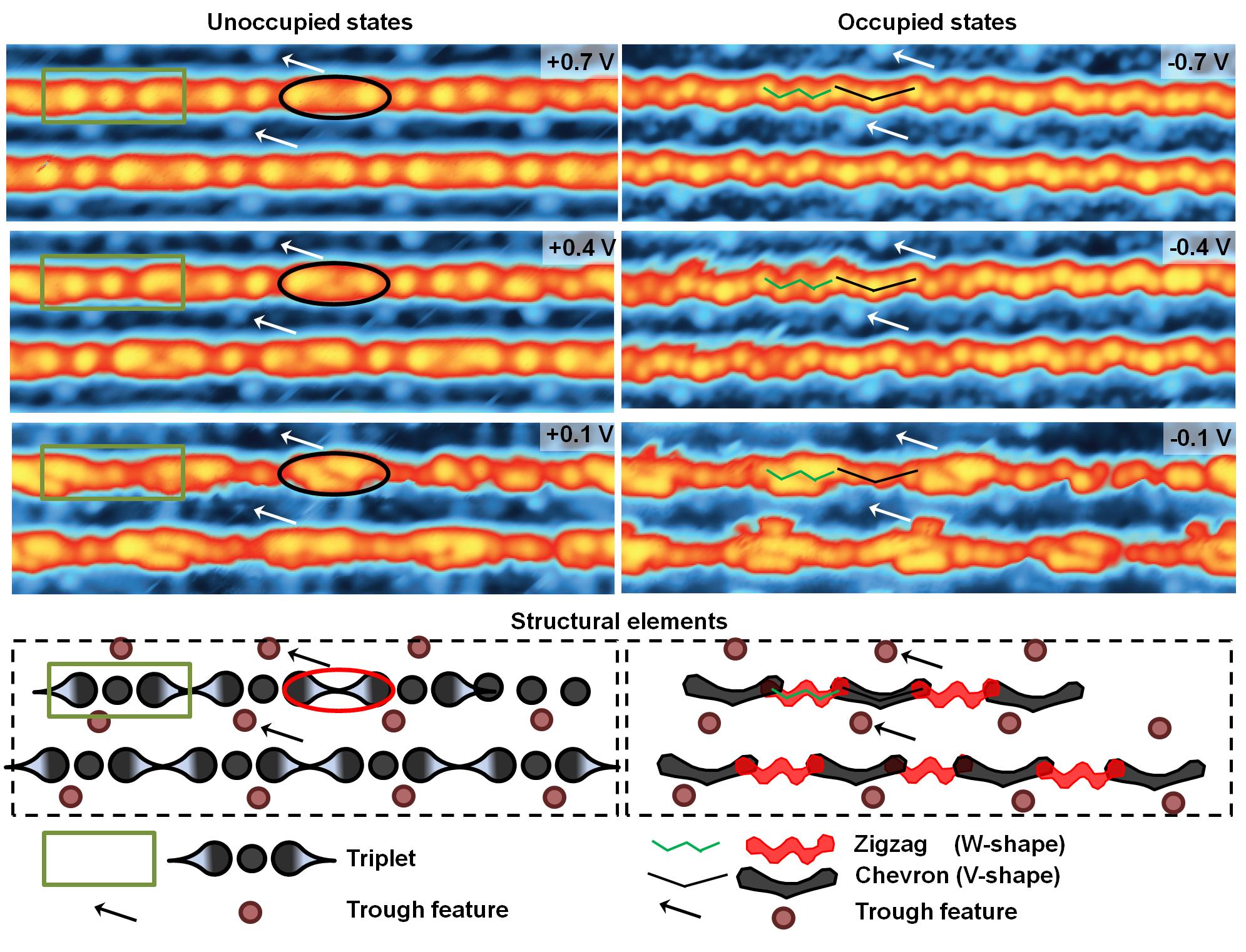}\\
  \caption{Low temperature STM images of Au NWs at various sampling biases, taken from Ref.~\onlinecite{BlumensteinC:JPhysCondensMatter2013}. Basic features are indicated. Unoccupied states: triplet feature on the NW is indicated with a green rectangle, white arrows indicate the trough feature between the NWs, and the overlap between the tails of the outer features of a triplet is indicated with a black ellipse. Occupied states: the chevron feature is indicated with a black V shape, while the zigzag feature is indicated by a green W shape. White arrows indicate the position of the trough features.}\label{fig:AuNW_chevron}
  \end{center}
\end{figure*}

\indent The experiments described in Sec.~\ref{ss:PtNWsExp} and \ref{ss:AuNWsExp} show that the single electron difference in the electronic configuration between Au and Pt matters greatly with regard to the resulting NWs. The experimental picture of Pt-induced NWs quickly converged, allowing for their use in various applied experimental setups,\cite{Oncel:ss06, Vriesde:apl2008, Kockmann:prb08, Saedi:nl09, KumarA:JPhysCondensMatter2012, HeimbuchR:PhysRevB2012}. Much progress has been made on the experimental picture of Au NWs too, but it remains one of controversy. Both the atomic structure and the electronic structure have been points of debate over the last decade, although the discussion in the (experimental) literature on the former seems to have settled in recent years. In the first experiments at room temperature, consistency between STM and LEED observations tended to be unclear, but it was found that at liquid nitrogen temperature the observations became clearer. Unfortunately for the theoretical modeler, matters also became much more complicated. Whereas the original observations of Wang \textit{et al.}\cite{WangJ:PhysRevB2004, WangJ:SurfSci2005} hinted at a ($4\times 2$) reconstruction, new observations by Sch\"{a}fer \textit{et al.}\cite{Schafer:prl2008} suggested a $c(8\times 2)$ reconstruction. Moreover, observations of van Houselt \textit{et al.}\cite{vanHouseltA:PhysRevB2008} suggested the NW height to be multiple stepsizes, in contrast to previous findings. However, low temperature STM experiments are required to observe a detailed structure structure of the Au NWs. At $77$ K a superstructure of the NWs was discovered, resulting in a $\times 4$ periodicity along the NW, or a periodic length of $32$ \AA.\cite{KockmannD:JPhysChemC2009, NiikuraR:PhysRevB2011, BlumensteinC:PhysRevLett2011} In addition to a more detailed structure of the NWs themselves, also a periodically reappearing protrusion is observed in the troughs between the NWs, as can be seen in Fig.~\ref{fig:AuNW_STM}d. The fact that these protrusions are observed for different biases (\textit{cf.}~Fig.~\ref{fig:AuNW_chevron}) and with different STM tips, shows them not to be mere tip artifacts.\cite{BlumensteinC:JPhysCondensMatter2013} On top of the NWs, two types of structures are visible in occupied state images: ``zigzag'' and ``chevron'' structures, shown in Fig.~\ref{fig:AuNW_chevron}. The trough protrusions seem to be exclusively linked to the latter, being located either left or right of the center peak. At this date, experimental observations seem to have more or less converged on this STM (atomic) structure.\\
\indent For the electronic structure and, more specifically, the possible presence of a $1$D metallic band along the NWs, the debate is still hot and glowing.\cite{Schafer:prl2008, vanHouseltA:PhysRevLett2009Comm, SchaferJ:PhysRevLett2009Reply, BlumensteinC:NatPhys2011, NakatsujiK:natphys2012, BlumensteinC:natphys2012Reply, HeimbuchR:NaturePhys2012} To resolve this point, a series of ARPES experiments have been published since $2009$, which show the nature of this band to be a complicated issue, in which experimental limitations often preclude a definite and final answer.\cite{NakatsujiK:PhysRevB2009, NakatsujiK:PhysRevB2011, MeyerS:PhysRevB2011, BlumensteinC:JPhysCondensMatter2013} The experiments show that there exists a highly anisotropic electronic band near the Fermi-level, which may well be 1D. The presence of a second-order phase transition at $585$ K, which appears to be of a 3D Ising type, indicates the NWs to be influenced by the substrate.\cite{BlumensteinC:PhysRevLett2011} On the other hand, spatial d$I$/d$V$ maps seem to support the $1$D picture by showing the conductance to be along the NW direction.\cite{HeimbuchR:NaturePhys2012, BlumensteinC:JPhysCondensMatter2013} Similar as for the Pt NWs, Heimbuch \textit{et al.}\cite{HeimbuchR:NaturePhys2012} show the highest differential conductivity to be located in between the NWs.\\
\indent These controversies make theoretical modeling far from trivial. Ironically, quite often similar controversies are solved by investigation of the atomistic models of the system, leading to a chicken and egg type problem in this case.

\begin{figure*}[!t]
  \includegraphics[width=16cm,keepaspectratio]{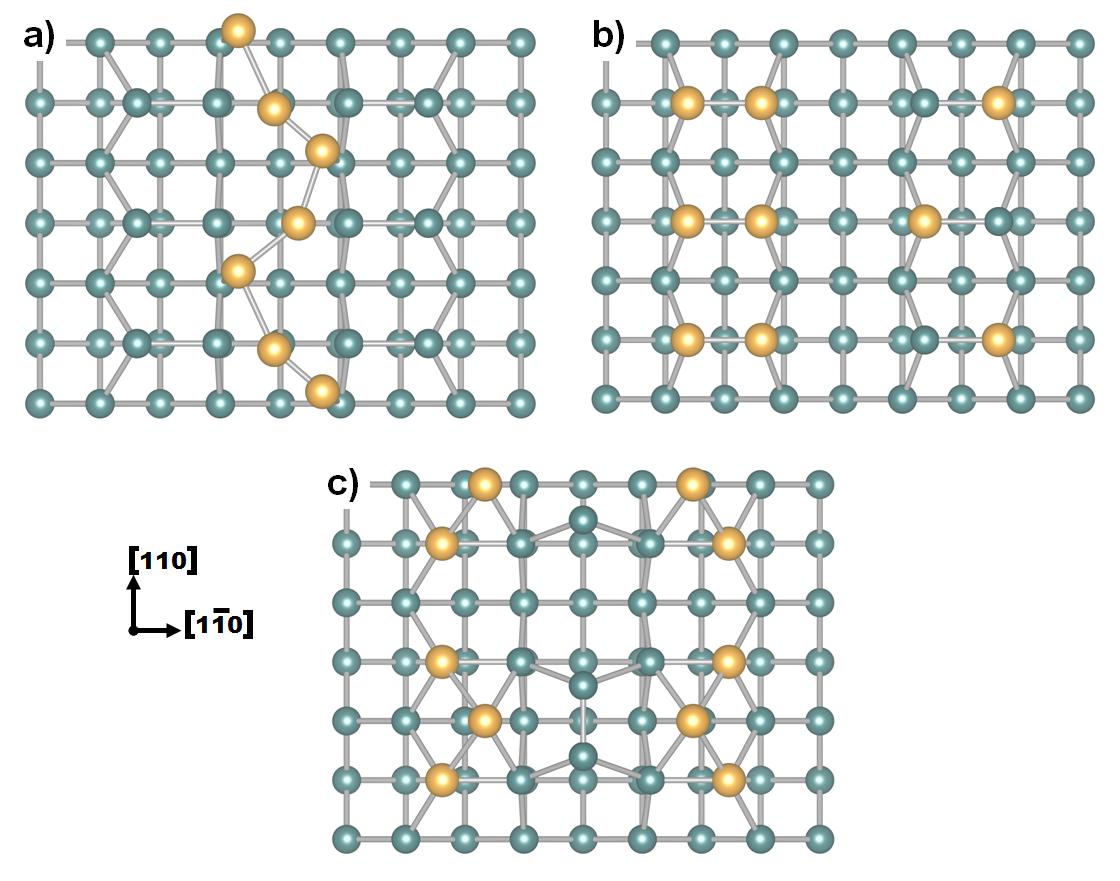}\\
  \caption{Ball-and-stick models of a representative structure of the classes of simple Au NW models. (a) Sch\"{a}fer class model: a zigzag Au chain with $4$ atoms periodicity located in the trough between the dimer rows of a reconstructed Ge(001) surface. This is the GC3 model presented in Ref.~\onlinecite{SauerS:PhysRevB2010}. (b) The model proposed by Wang \textit{et al.}\cite{WangJ:PhysRevB2004}(WM), containing alternating Au-dimer rows and mixed dimer rows. (c) The EBD model, which was the most stable Sauer Bridging Dimer Class Models, showing Ge-dimers forming the NW, imbedded in the Ge-lined troughs of a mixed dimer reconstructed Ge(001) surface. Additional Au atoms on the QDRs stabilize the structure, and are proposed as the structural elements which give rise to the features in the troughs between the NWs observed in low temperature STM images.}\label{fig:AuNWmodels}
\end{figure*}
\begin{figure*}[!t]
  \includegraphics[width=12cm,keepaspectratio]{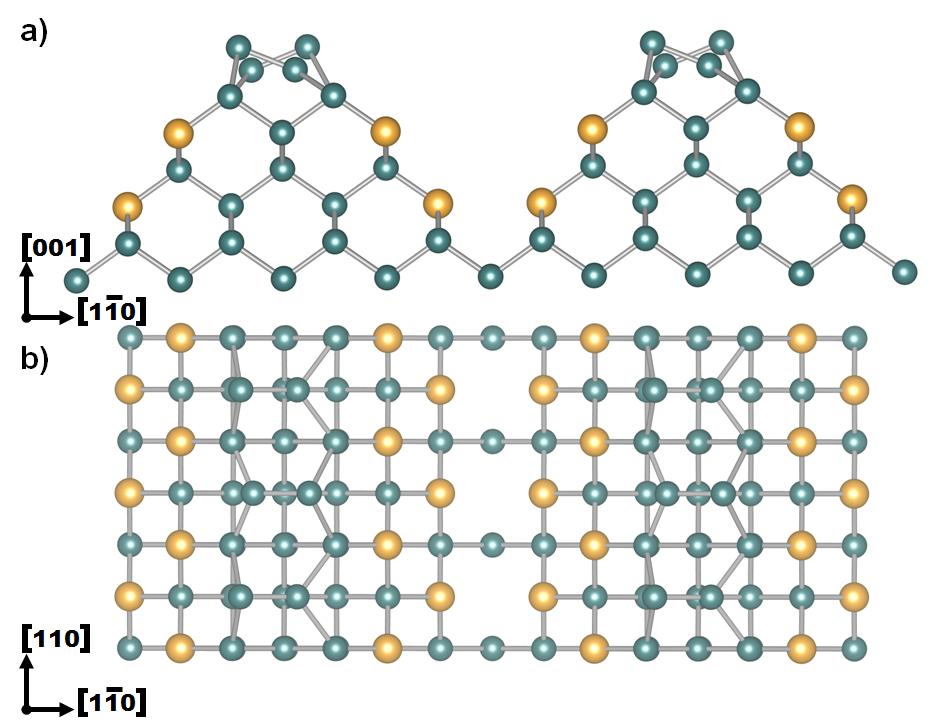}\\
  \caption{Ball-and-stick model of the Giant Missing Row (GMR) model proposed by van Houselt \textit{et al.}\cite{vanHouseltA:PhysRevB2008}, after the model presented by Sauer \textit{et al.}\cite{SauerS:PhysRevB2010}: (a) side view, (b) top view. }\label{fig:AuNWmodels_GMR}
\end{figure*}

\subsubsection{Four classes of models}
\indent In contrast to the Pt NW system, and the experimental literature on the Au NWs, there are only a few theoretical studies covering the Au NWs.\cite{LopezMorenoS:PhysRevB2010, SauerS:PhysRevB2010, MeyerS:PhysRevB2012} In these works, a wide range of Au concentrations ($0.25$ to $1.00$ ML) and a large variety of different surface reconstructions have been investigated, including the models proposed by experimentalists. These models can be split into four classes presented below:\cite{fn:AuModel}
\begin{description}
  \item[Sch\"{a}fer Class Models (SCMs)] In this class, all models consist of a $c(4\times 2)$ or a $b(2\times 1)$ reconstructed Ge(001) substrate, with adsorbed Au atoms or dimers. It is based on the work of Sch\"{a}fer \textit{et al.}\cite{Schafer:prl2008}, in which these authors proposed the NW to be a single Au atom wide, suggesting the NW to consist Au chains. In this class, the Au atoms form linear or zigzag chains, either on the Ge-dimer rows or in the troughs in between. An example is shown in Fig.~\ref{fig:AuNWmodels}.
  \item[Wang Class Models (WCMs)] This class of models contains the originally proposed model by Wang \textit{et al.}\cite{WangJ:PhysRevB2004}, in which surface dimers of the pristine Ge(001) surface are replaced by either Ge--Au heterodimers or Au homodimers. All models with Ge--Au and/or Au--Au dimers in the top layer (only) are considered in this class. The model proposed by Wang \textit{et al.} (WM) is shown in Fig.~\ref{fig:AuNWmodels}.
  \item[Sauer Bridging Dimer Class Models (SBDCMs)] In this third class of models, suggested by Sauer \textit{et al.}\cite{SauerS:PhysRevB2010}, a Ge(001) surface modified with $0.5$ ML of Au is extended with the addition of Au- or Ge-dimers. The Au-modified Ge surface consists of Au--Ge heterodimers ordered in such a way as to alternatingly result an Au-lined trough and a Ge-lined trough. The additional Au- or Ge-dimers are placed in the Ge-lined trough (the opposite of similar Pt NW models\cite{VanpouckeDannyEP:2008PhysRevB, VanpouckeDannyEP:2009MaterResSocSympProc_Nanowire, VanpouckeDannyEP:2010bPhysRevB_NanowireLong}). In addition, isolated Au atoms can also be present on the QDRs; an example of such a structure is shown in Fig.~\ref{fig:AuNWmodels}.
  \item[van Housel Class Models (HCMs)] The fourth class consists of models containing complex reconstructions, in which large variations in height are present. In these models, the remaining dimer rows are flanked by Ge(111) facets, which in turn are decorated with Au atoms or trimers. The reconstruction of the top layer of these NWs varies greatly from model to model. These models originate from the work of van Houselt \textit{et al.}\cite{vanHouseltA:PhysRevB2008}, who noted corrugations of at least $6$ \AA\ for the Au NW. The Giant Missing Row (GMR) model suggested by these authors is shown in Fig.~\ref{fig:AuNWmodels_GMR}.
\end{description}
\subsubsection{Simple models for Au nanowires}
\indent Starting from a clean substrate, the initial construction of a NW is quite often considered in terms of adsorbed atoms or dimers. Theoretical studies on the adsorption of Au-dimers on the Ge(001) surface show Au-dimers to be most stable in the troughs parallel to the dimer rows.\cite{LopezMorenoS:PhysRevB2010, NiuCY:SolStateComm2011} Extending the Au-dimers to chains does not change this trend. However, this type of structures was also shown to have unfavorable energetics excluding the SCMs from the search for Au NWs.\cite{SauerS:PhysRevB2010} Imbedding the Au atoms into the top layer of the substrate improves the situation, in some cases even resulting in reconstructions which are more stable than the original Ge(001) surface (depending on the functional used).\cite{SauerS:PhysRevB2010} For $0.25$ ML of Au imbedded as heterodimers in the top layer, both Sauer \textit{et al.} and L\'{o}pez-Moreno \textit{et al.} find the most stable structures to be identical to those found in case of Pt ($\beta_4$ and $\beta_6$ reconstructions).\cite{VanpouckeDannyEP:2010aPhysRevB_BetaTerrace, SauerS:PhysRevB2010, LopezMorenoS:PhysRevB2010} Just as for the Pt case, the Au--Ge heterodimers tend to flatten out. A comparison of the electronic band structures of the $\beta_{6u}$-reconstruction for Pt/Ge(001) (Fig.~8b in Ref.~\onlinecite{VanpouckeDannyEP:2010aPhysRevB_BetaTerrace}) to the $2$--$8$ structure for Au/Ge(001) (Fig.~3b in Ref.~\onlinecite{LopezMorenoS:PhysRevB2010}) show an almost identical picture. Bands near the Fermi level show a large dispersion along the $\Gamma$--$J$ and $K$--$J^{\prime}$ directions (parallel to the NW), while there is nearly no dispersion along the $J$--$K$ and $J^{\prime}$--$\Gamma$ directions (perpendicular to the NW). This appears to be in agreement with the experimental observation of a 1D band in ARPES measurements by Sch\"{a}fer and co-workers.\cite{Schafer:prl2008, MeyerS:PhysRevB2011}\\
\indent Similar as for the Pt $\beta$-terraces, Sauer \textit{et al.}\cite{SauerS:PhysRevB2010} and L\'{o}pez-Moreno \textit{et al.}\cite{LopezMorenoS:PhysRevB2010} also find that in simulated STM images, the Au atoms do not present themselves as bright spots, but rather as darkened regions. As such, one might expect to find $\beta$-terraces on Au-deposited Ge(001) surfaces, similar to the ones found for the Pt-covered substrates.\cite{Gurlu:apl03, VanpouckeDannyEP:2010aPhysRevB_BetaTerrace, SauerS:PhysRevB2010, LopezMorenoS:PhysRevB2010} However, unlike its Pt counterpart, Au-dimers imbedded in the Ge(001) toplayer are much more stable than heterodimers.\cite{fn:AuDimers, VanpouckeDannyEP:2010aPhysRevB_BetaTerrace, SauerS:PhysRevB2010, KumarA:PhysBCondMatter2011} It was even shown that for $1$ ML Au deposition replacing all surface Ge-dimers by Au-dimers leads to a very stable configuration, which agrees very well with the experimental observation of the formation of 3D Au islands under high deposition amounts.\cite{SauerS:PhysRevB2010, PopescuDG:PhysStatusSolidiRRL2013} It also explains why no $\beta$-terraces are observed for the Au-deposited systems.\\
\indent Since the Au NW model originally suggested by Wang \textit{et al.}\cite{WangJ:PhysRevB2004} consists only of Ge--Au heterodimers and Au--Au homodimers, it is to be expected that such a reconstruction is relatively stable, as was shown by Sauer \textit{et al.}\cite{SauerS:PhysRevB2010}, albeit less stable than the fully Au-covered surface. In this model, $0.75$ ML of Au is included in the top layer, putting it in very good agreement with the most accurate experimental determination of ($0.75\pm0.05$) ML Au coverage by Gallagher \textit{et al.}\cite{GallagherMC:PhysRevB2011} In simulated STM images bright NWs are visible. These consist of the QDR built of Au--Ge dimers, where the Ge atoms give rise to the NW image. The troughs between these zigzag chains consist of a dimer row built of flat Au--Au homodimers.
Although this is the exact opposite of the interpretation by Wang \textit{et al.}\cite{WangJ:PhysRevB2004}, the simulated STM images show this model to be in reasonable agreement with those early experiments. Unfortunately, the simulated STM images of this model cannot explain the features observed in low temperature STM studies of the Au NWs, nor the large corrugations and depths of the troughs between the NWs as observed by van Houselt \textit{et al.}\cite{vanHouseltA:PhysRevB2008, KockmannD:JPhysChemC2009, NiikuraR:PhysRevB2011, BlumensteinC:PhysRevLett2011, BlumensteinC:JPhysCondensMatter2013, SafaeiA:PhysRevB2013}\\
\indent Recently, the WM was revisited by Meyer \textit{et al.}\cite{MeyerS:PhysRevB2012} in a combined surface-XRD/DFT study. Surface-XRD data only yields the intensity of the structure factor $F_{hkl}$ (i.e. $|F_{hkl}|^{2}$) and thus prevents the direct calculation of the electron density through Fourier transform.Therefore, Meyer \textit{et al.}\cite{MeyerS:PhysRevB2012} make use of a Patterson map. The Patterson function $P(\mathbf{r})$ is given by:\cite{PattersonAL:PhysRev1934, PattersonAL:ZKristallogr1935}
\begin{equation}
P(\mathbf{r}) = \int{\rho(\mathbf{r^{\prime}})\rho(\mathbf{r^{\prime}}+\mathbf{r})d\mathbf{r^{\prime}}} =\sum_{hkl}{|F_{hkl}|^{2}e^{-i\mathbf{q}\centerdot\mathbf{r}}},
\end{equation}
which is the autocorrelation function of the electron density $\rho(\mathbf{r})$ obtained by applying a Fourier transform to the intensities of the structure factors. Since interatomic distances are present in the electron density, they will also be present in the Patterson function, allowing one to obtain lengths and directions between surface atoms, although not absolute positions. In a Patterson map, the intensity of the peak scales with the product of the atomic numbers of the contributing atoms. As such, the highest peaks are attributed to Au--Au distances, while the second and third highest peaks are associated with Au--Ge and Ge--Ge distances, respectively. Using such a map, Meyer \textit{et al.}\cite{MeyerS:PhysRevB2012} built a \textit{minimum structural model} which turned out the be the original WM. However, upon relaxation of this model using \textit{ab-initio} DFT calculations, it was found that the resulting buckling of the heterodimers gave rise to a splitting of the peaks in the associated Patterson map. This led the authors to conclude that this model already contains some good elements, but further refinement is needed.\\
\subsubsection{Bridging dimers or missing dimer rows}
\indent The models of the previous section taught us some important lessons: (1) Au imbedded in a Ge(001) substrate is invisible for STM, (2) well ordered Au--Ge heterodimers improve the stability of the  substrate, (3) the large depth of the troughs between the NWs cannot be explained by the WM, (4) nor can it explain the low temperature STM features.\\
\indent Based on the experimental suggestion that deeper troughs are required, Sauer \textit{et al.}\cite{SauerS:PhysRevB2010} investigated a set of reconstructions containing dimers bridging every second trough of the Ge(001) surface. In the most successful reconstructions, the substrate surface consisted of Ge--Au mixed dimers only,\cite{SauerS:PhysRevB2010} in a configuration dubbed the $\gamma_{\mathrm{as}}$ reconstruction in case of the Pt/Ge(001) system.\cite{VanpouckeDannyEP:2008PhysRevB, VanpouckeDannyEP:2010bPhysRevB_NanowireLong} This surface reconstruction consists of QDRs with troughs, which are alternatingly lined with Ge and metal atoms (\textit{cf.}~Fig.~\ref{fig:AuNWmodels}c). Unlike the Pt case, Sauer \textit{et al.}\cite{SauerS:PhysRevB2010} places the NW dimers in the Ge-lined trough. Such a reconstruction provides [$1\overline{1}4$] and [$\overline{1}14$] facets, which lead to stable Ge surfaces. In contrast to Au-dimers adsorbed on a pure Ge surface, this reconstruction is quite stable ($-65$ meV/($1\times 1$)\cite{fn:GGAenergy} unit cell). This is slightly more stable than when a Ge bridging dimer is used ($-50$ meV/($1\times 1$) unit cell). However, the latter can be stabilized further (to $-82.5$ meV/($1\times 1$) unit cell) by the addition of isolated Au atoms on the QDRs, and is shown in Fig.~\ref{fig:AuNWmodels}. Unfortunately, at this point it is important to note that these formation energies are less favorable than those for the WM ($-87.5$ meV/($1\times 1$) unit cell) or a surface fully covered in Au-dimers ($-152.5$ meV/($1\times 1$) unit cell). The resulting STM images are very rich in shapes and features, and show individual dimers and atoms. These last aspects are clearly not in agreement with the experimental STM images. Sauer \textit{et al.}\cite{SauerS:PhysRevB2010} also investigated a TDC model for the Au NW, but found this to be energetically very unfavorable ($+86.25$ meV/($1\times 1$) unit cell). So, it would appear that different models which give good hints toward the atomic structure of Pt NWs are totally off when it comes to Au NWs.\\
\indent Following the suggestion of van Houselt \textit{et al.}\cite{vanHouseltA:PhysRevB2008}, Sauer \textit{et al.} also investigated the GMR reconstruction, shown in Fig.~\ref{fig:AuNWmodels_GMR}. Although the simulated STM images show good agreement with the experimental high temperature STM images, and the GMR structure is metallic, the obtained formation energy is highly unfavorable ($+185$ meV/($1\times 1$) unit cell). Furthermore, this reconstruction would require $1.0$ ML of deposited Au, in contrast to the results obtained by Gallagher \textit{et al.}\cite{GallagherMC:PhysRevB2011} Also, after relaxation, the Au atoms do not form trimers, but just build into the bulk structure of the Ge lattice. Modifying the GMR reconstruction to some extent, Sauer \textit{et al.}\cite{SauerS:PhysRevB2010} were able to stabilize the structure ($-55$ meV/($1\times 1$) unit cell). This Au-trimer stabilized Ge-ridge (ATSGR) structure contains a linear chain of Ge atoms at its apex, with Au-trimers on the faces of the ridge. The resulting electronic band structure still shows it to be a metallic system. Moreover, metallic bands with an almost 1D character occur along the direction of the NW. However, the specifics of these bands do not appear to be in agreement with ARPES findings. The simulated STM images show a clear linear NW with very little features (independent of bias), in reasonable agreement with the high temperature STM images.\\
\subsection{Conclusions and outlook on Au nanowires}
\indent Although a very large number of structures, with a wide range of Au concentrations, have already been investigated in the theoretical literature, the authors have not been able to put forward a definite structure which is both energetically favorable and results in simulated STM images in agreement with all the experimental findings. Theoretical work also showed the experimentally suggested structures to be either energetically unfavorable or not to result in the observed STM features. Despite this, the findings of these works do provide useful pointers for future investigations. Just as for the Pt system, Au atoms imbedded in the top layer of Ge(001) are invisible for STM, hinting that the observed NWs may consist of Ge atoms. Bridging dimer models, which are successful for the Pt NWs, do lead to stable structures with both Ge- and Au-dimers, but the associated simulated STM images do not show the features observed in low temperature STM experiments. These structures do show better agreement with the experimentally observed very large trough depths. Also the HCMs show this large trough depth, but the respective simulated STM images only show agreement with the room temperature STM images, indicating a more detailed structure must be present in the real Au NWs than is present in the current models. Since different models are successful at explaining different experimental observations, further model design will have to aim at combining into a single model the parts leading to successful features.\\
\indent Although the Au coverage for the Au NWs is accepted to be $0.75$ ML, further experimental clues will be required before theoretical modeling may provide a breakthrough. ARPES measurements have already provided valuable information on the electronic structure, which is ideal for testing models, but additional hints on the atomic structure are missing. Detailed studies of Patterson maps of different models should lead to new insights. The possible presence of metallic bands near the Fermi level, and the fact that going from room temperature down to $77$ K leads to a significant increase in detail on the Au NWs, would suggest that going to lower temperatures still may provide even more structural information. Also, further detailed spatial mapping of the conducting states,\cite{HeimbuchR:NaturePhys2012} indicating the states to be present between or on top of the wires, will provide insights for theoretical models, similarly as they did for the Pt NWs (\textit{i.e.} Ge NWs imbedded in Pt-lined troughs). Due to the reasonable success when revisiting the WM with Patterson mapping, the question of the actual depth of the troughs becomes an important point again. Is the origin of this height variation electronic or geometric in nature? In the former case, simple structures will regain in importance.\\
\indent In addition, alternate approaches of probing the NWs should be considered. Recall that the Pt NWs were decorated with CO molecules to test their nature and that the presence of asymmetric STM images of the molecules supported the PINW models; similar investigations are possible for the Au NWs. Another interesting experiment would be to try to pick up part of the NW with an STM tip and remove it, to allow one to peek underneath, just as was done in the Pt case.\\
\indent Finally, Au NWs can show interdigitated regions unlike the Pt NWs, which hints at a basic unit that can be less than $16$ \AA\ wide. Detailed STM studies of these regions will yield invaluable information for model builders. More generally, STM images showing perfect terraces filled with NWs are ideally suited for publications and front covers, but it are the defects, step edges, domain boundaries and other regions where things lead to ``\textit{ugly}'' STM images, that contain crucial information for model design.\\
\section{Other nanowires on Ge(001)}
\indent In the previous sections, it was shown that the anneal temperature plays an important role in the formation of NWs. For the Pt/Ge(001) system, no NWs are observed for anneal temperatures below $1000$K, while for the Au/Ge(001) system a limited temperature window for NW growth is found. Studying the Pt--Ge phase-diagram, one quickly observes that the anneal temperature of $1050$ K lies roughly on top of the phase-boundary between solid and (solid +) liquid (L) phases. More specifically, the reaction $\mathrm{L} \leftrightarrow \mathrm{Ge_{bulk}} + \mathrm{Ge_{2}Pt}$ is found at a temperature of $770^{\circ}$C ($= 1043$ K).\cite{Massalski:BAPD90} Similarly, the In NWs observed by Falkenberg \textit{et al.}\cite{Falkenberg:ss1997,Falkenberg:prb2002} and more recently the Au NWs appear at a sample temperature just above the solid/solid+liquid phase-boundary. In each case, a substantial rearrangement of the surface atoms takes place, which means that bonds of the surface atoms need to be broken. To achieve this bond breaking, temperatures close to the melting temperature are required.\\
\indent The examples provided above give a good indication that even more metal/Ge(001) NW systems should exist, each with their own unique properties. One could imagine magnetic NW systems using Co, where the $\mathrm{L} \leftrightarrow \mathrm{Ge_{bulk}} + \mathrm{CoGe}$ reaction at $817^{\circ}$C ($=1090$ K) gives an estimate for the required anneal temperature.\cite{VanpouckeDannyEP:2009ThesisNW, MuzychenkoDA:PhysRevB2012} In $2011$, it was shown that $1$D wires are formed for $0.1$ ML Co deposited on Ge(001) annealed at $700$ K.\cite{ZandvlietHJW:SurfSci2011} With a width of $16$ \AA, these NWs are quite wide compared to the Au and Pt NWs. However, the anneal temperature used is about $400$ K below the suggested $1090$ K anneal temperature, so other NW phases with thinner NWs may exist for higher anneal temperatures.\\
\indent Metals like Cu and Ni have a similar negligible solubility in Ge while showing interesting reactions in their phase-diagrams, resulting in anneal temperatures of $644^{\circ}$C ($=917$ K) for Cu/Ge(001) and $762^{\circ}$C ($=1035$ K) for Ni/Ge(001).\cite{Massalski:BAPD90} \\
\indent Chain formation of Pt and Au on Ge(001) is often linked to $sd$ competition due to the relativistic character of the $5d$ electrons, which would lead to a preference for low coordination.\cite{Takeuchi:prl89, Yanson:nat98, Smit:prl01, Gurlu:apl03, WangJ:PhysRevB2004, OncelN:JPhysCondensMatter2008} Based on this, NW formation is both predicted and observed for Ir,\cite{MockingTF:naturecomm2013} Pt,\cite{Gurlu:apl03} and Au.\cite{WangJ:PhysRevB2004} The observation Ir NWs on Ge was done only very recently, and showed behavior that is interesting for study within atomistic quantum mechanical frameworks. The confirmation of the claim that a standing wave is present in these Ir NWs would be of great interest. For the theoretical community, this would provide a new toy model to investigate quantum mechanics. Using atoms in molecules partitioning (AIM) schemes, it would be possible to check how the standing wave behavior is linked to charge transfer inside the NW. At the same time, it provides an interesting system to compare different AIM schemes leading to a better understanding of the very nature of atoms in a real world system.\cite{MullikenRS:1955aJChemPhys, MullikenRS:1955bJChemPhys, MullikenRS:1955cJChemPhys, MullikenRS:1955dJChemPhys, HirshfeldFL:1977TCA, BaderRFW:book1990, BaderRFW:1991ChemRev, LillestolenTCWheatleyRJ:2008ChemComm, LillestolenTCWheatleyRJ:2009JChemPhys, VanpouckeDannyEP:2013aJComputChem, VanpouckeDannyEP:2013bJComputChem} For the modeling community, on the other hand, this could be used to establish the validity of the presented model of the Ir NW, which in turn would provide guidelines for the experimental community allowing them to investigate the system further, for example for its use in molecular electronics applications.\\
\indent Also for other sixth row elements, NW formation has been observed, including elements without filled $5d$ shell: \textit{e.g.} Ba/Ge(001),\cite{LukanovBR:PhysRevB2011} Ho/Ge(111),\cite{Eames:prb06} and Er/Ge(111).\cite{Pelletier:JVS2000}\\
\indent Despite the existence of these experimental reports, only very few theoretical works address NWs on Ge, other than the Au and Pt NWs. The existing studies focus on one or a few models, which have been proposed in an \textit{ad hoc} fashion. Systematic studies would be of great interest for materials design applications. Their current absence may be due to the unit cell size of the systems: the observed surface reconstructions require quite large surface cells, which makes them relatively expensive. In the following two paragraphs, we take a brief look at a few systems for which some theoretical work is already available.
\paragraph{Co on Ge}
\indent Muzychenko \textit{et al.}\cite{MuzychenkoDA:PhysRevB2012} have studied the initial adsorption of single Co atoms on the Ge(111) ($2\times 1$) surface. In this combined theoretical/experimental study, it is found that the Co atoms move into the substrate, \textit{i.e.} into the large seven-member Ge ring. As a result, the dimers above the Co atom light up brightly in STM for large positive bias. Although the deposition amount of $2$--$4$ \% ML is too low for NW formation, from these results one may expect that a NW reconstruction is not unlikely to present itself at deposition amounts comparable to those for Pt and Au NWs.\cite{MuzychenkoDA:PhysRevB2012} Additionally, recent experiments by Zandvliet \textit{et al.}\cite{ZandvlietHJW:SurfSci2011} show that submonolayer deposition of Co on Ge(001) leads to hexagonal islands at low temperatures ($550$ K), while higher temperatures give rise to NWs. The size of these structures, however, will make theoretical modeling rather difficult: the hexagonal islands each are $3\times 4$ surface dimers or $48$ surface atoms in size. Simulation of such a structure would require one to have several hundred atoms in a unit cell, precluding anything but the most basic calculations.\\
\paragraph{Sr on Ge}
\indent Deposition of alkali-earth elements such as Ba and Sr at elevated temperature roughens the Ge(001) surface. With increasing Sr coverage, several ordered surface reconstructions are observed and highly ordered arrays of 1D stripes can be produced.\cite{LukanovBR:PhysRevB2011} Just as for the Co case, only a combined experimental/theoretical study at low Sr coverage is available at this time. Using simulated STM images for \textit{ab-initio} calculated minimum energy structures, Lukanov \textit{et al.}\cite{LukanovBR:PhysRevB2011} are able to identify observed Sr-induced reconstructions at $1/6$ ML coverage. Chainlike structures of Sr atoms located in the troughs between the Ge-dimer rows are obtained. Since Sr incorporation is found to eject Ge atoms onto the terraces, also Ge-dimers are included in the surface reconstructions, resulting in troughs which are filled with Sr atoms and Ge-dimers in an alternating fashion. Similar as for the Au and Pt NWs, it is found that the Sr atoms donate valence electrons to Ge, leading the latter to light up brightly in STM.
%------------------------------------------------------------------------------------------
%----------------------------Conclusions---------------------------------------------------
%------------------------------------------------------------------------------------------
\section{Outlook: Metal-induced nanowires on
semiconductor surfaces and phase-space searching}
\indent Modern technology is driven by the constant further miniaturization of devices. With the scale of these devices reaching deeper and deeper into the microscopic regime, standard fabrication methods are quickly becoming too crude to build these (near-)atomic scale devices. Self-assembly is pursued actively as an alternative for the current lithographic techniques; \textit{i.e.}\ let nature do the hard work.\cite{Barth:nat05} However, before it becomes possible to actively design devices at an atomic scale, a large database of possible different components will be required, containing an accurate description of their properties and structure and all the necessary conditions. To design such components, a symbiotic collaboration between experimental and theoretical researchers will not just be an asset, it will become a \textit{condicion sine que non} to succeed. This necessity follows from the nature of this subject. The limitations on experiment and theory prohibit each individually to have a full understanding.\\
\indent On the one hand, the experimentalist can build new structures, but (s)he can never be absolutely certain of the exact atomic structure. This forces him/her to make assumptions, which greatly influence the interpretation of the observations. (For example, compare the original experimental interpretation of the nature of the Pt-induced NWs, and the adsorption of CO onto them, to the later theoretical explanations). Close collaboration with theoretical researchers can illuminate the situation. Through \textit{ab-initio} calculations and direct comparison, the experimentally observed structures can be identified, making the interpretation of the observations less of a guessing game.\\
\indent On the other hand, since the structures under study are mostly metastable configurations, the theoretician  will be at a loss without an experimental reference. The phase-space of possible structures he/she is working with is gigantic, and the number of local minima is large. Identifying which of these minima correspond to ``real'' (metastable) experimental configurations can only be done through direct comparison to experiments. Furthermore, since the most commonly used methods are ground state methods, they are ideally suited to find the ground state, \textit{i.e.}\ global minimum, of a system. However, when dealing with metastable configurations the ground state is not the configuration sought after. Since metastable configurations show a strong dependence on the experimental conditions (\textit{e.g.}\ deposition rate and amount, temperature, \ldots), accurate and complete experimental information is necessary to reduce the phase-space of possible structures to a surveyable size for the theoretician. Once the structure is identified successfully, theoretical work can be used to fine-tune the experimental parameters, or even predict other closely related metastable configurations.\\
\indent In both directions, from theory to experiment and from experiment to theory, direct comparison plays an important role. With direct comparison we mean that as few layers of modeling (and assumptions) should be present between the data from the experimental observation and the data from theoretical calculations. STM is a magnificent tool in this regard, as it shows almost pure data with regard to the system under study. Ignoring tip effects (which can often be identified quickly, \textit{cf.}\ the Au NWs) STM is generally speaking a \textit{what-you-see-is-what-you-get} approach. In comparison, obtaining crystallographic data (atomic positions and lattice vectors) from XRD is an indirect approach requiring prior knowledge of the structure, and fitting of the experimental data to these assumptions, which in turn can easily lead to contradictory observations and results. Although Patterson mapping requires XRD data, it may be considered another (almost) ``direct'' comparison approach, since the mathematical operations needed on theoretical and experimental data are limited and equivalent; in particular no fitting is required.\\
\indent It needs to be noticed, however, that STM and Patterson map comparison are rather qualitative instead of quantitative, adding to their robustness. Unlike energies and lattice parameters, they do not allow for an (easy) quantification of the difference between results in a single simple meaningful number.\cite{WenmackersS:StatNeer2012}\\
\ \\
\indent Let us now return to the case at hand: nanowires on Ge.
The formation of nanowires, chains, and rods has been observed on Ge surfaces
after deposition of Pt, \cite{Gurlu:apl03, Gurlu:prb04, Oncel:prl05, Oncel:ss06, Schafer:prb06, Fischer:prb07, Houselt:nanol06, Houselt:ss08, Kockmann:prb08, ZandvlietHJW:JPhysCondMatter2009, MochizukiI:PhysRevB2012, HeimbuchR:JPhysCondensMatter2013} Au,\cite{WangJ:PhysRevB2004, WangJ:SurfSci2005, Schafer:prl2008, vanHouseltA:JPhysCondensMatter2010, MockingTF:SurfSci2010, MeyerS:PhysRevB2011, NiikuraR:PhysRevB2011, MeyerS:PhysRevB2012, MelnikS:SurfSci2012, BlumensteinC:JPhysCondensMatter2013} In,\cite{Rich:prb1990, Falkenberg:ss1997, Falkenberg:prb2002, ZHQin:CPB2008} Er,\cite{Pelletier:JVS2000} Ho,\cite{Eames:prb06} Co,\cite{ZandvlietHJW:SurfSci2011} Ir,\cite{MockingTF:naturecomm2013} Yb,\cite{KuzminM:SurfSci2013} Sr,\cite{LukanovBR:PhysRevB2011, LukanovB:PhysRevB2012} and Ba.\cite{LukanovBR:PhysRevB2011} Many of these systems show a large variety of surface reconstructions at different submonolayer depositions and temperatures (\textit{cf.}, the surface phase-diagram for In on Ge(001) given in Ref.~\onlinecite{Seehofer:ss1996}). The fact that `nanowires' have been observed for each of these combinations shows that this type of reconstruction is not a rare exception, highly dependent on the constituent atomic species. It does show, however, that the reconstructed Ge(001) is an ideal surface for what is called \textit{template-driven} nanowire formation. The successful formation of $1$D structures on reconstructed Ge and Si(001), and on high-index surfaces of Si, show that this type of surface reconstructions can be used in the quest for smaller electronic devices. The study of Pt and Au NWs also shows that the obtained NWs can show very different behavior, allowing the NWs to be tailored for specific needs by simply using other elements.\\
\indent Based on the studies presented in this review, some important conclusions for designing NWs can be formulated:\\
\begin{itemize}
  \item Surface reconstructions (dimer rows, step edges, but also threefold symmetric Ge(111) reconstructions\cite{Eames:prb06}) provide $1$D templates facilitating the formation of NWs that can be hundreds of angstroms long.
  \item To preserve the semiconducting nature of the substrate, it is
beneficial if the ``metal'' atoms have a (nearly) zero solubility in the
semiconductor. This has the additional advantage that the amount
deposited will be exactly the amount needed for the NW
reconstruction, giving a better control over the process.
  \item The NW reconstruction aimed for can be considered extreme cases of interface systems. As such, the bulk diffusion should be negligible and submonolayer deposition should suffice in most cases. This will also benefit the decoupling of the NWs and the substrate.
  \item Surface mobility and mass-transport play a crucial role during the formation of 1D and 2D surface structures. This can be deduced from the formation of Au NWs spanning multiple terraces of the Ge substrate, and the distinctly different structures observed in the Er/Ge(001) system when a sample is only annealed for a short period of time, compared to deposition at an elevated temperature.\cite{Pelletier:JVS2000} In the latter case, the short annealing time has the advantage that bulk diffusion is limited.
  \item On the Ge(001) substrate, deposited metal atoms tend to be invisible in STM. As such, the observed NWs quite often consist of Ge atoms.
\end{itemize}

\indent Although this information is very useful for practical application, it also shows that simple one-sided theoretical prediction is impossible: each metal species has its own specific electronic structure and resulting binding behavior. Investigation of the Pt--Pt, Ge--Ge and Pt--Ge free dimers, shows the Pt--Ge dimer to be the most stable of all, while for the Au--Au, Ge--Ge and Au--Ge free dimers, the Ge--Ge dimer is found to be the most stable, followed by the Au--Ge dimer. The geometry of a successfully identified NW cannot simply be copied from one system to another. This is clearly demonstrated for the Pt and Au NWs.  Furthermore, ground state calculations are unable to give single structure predictions for metastable configurations; multiple structures will appear as possible solutions instead. Moreover, the actual experimental metastable configuration may even be missing. Due to the size of the phase-space, one cannot do an exhaustive search covering all possible configurations. In Ref.~\onlinecite{oganov:jcp2006} the number of possible starting configurations for a binary alloy containing $30$ atoms in the unit cell is estimated to be $10^{47}$; the number of inequivalent atoms in the unit cells presented in this review is easily $50$ or more. A clever search algorithm, such as a genetic algorithm (GA), can be used.\cite{oganov:jcp2006, Pickard:prl2006, Pickard:prb2007, Avezac:prb2008, PickardCJ:JPhysCondensMatter2011} GAs can cover the entire phase-space searching for specific properties, and have already been applied successfully for simple problems with relatively small supercells.\cite{oganov:jcp2006,Pickard:prl2006,Pickard:prb2007,PickardCJ:JPhysCondensMatter2011} Such GAs can be combined with \textit{ab-initio} calculations, and can thus be used to approach the current problem of nanowire-design. The average system size and the fact that a few hundred structures are often needed for the GA to converge to its solution makes this technique computationally expensive to use, but not impossible, unlike a simple brute force exhaustive search. Next to the computational cost, which at the moment of writing might be just too steep for common use in systems of the current size, there is also the inherent complexity of the ``fitness''-algorithm. Whereas magnetism and total energy are given in absolute numbers, the property `\textit{resembles a nanowire in STM}' is too vague for simple binary logic. For a human being, however, this problem is surmountable. In fact, if we look at the work presented in Ref.~\onlinecite{VanpouckeDannyEP:2010bPhysRevB_NanowireLong}, the method applied there for finding the Pt NW geometry can be considered a type of GA. Only in this case, it was the author who performed all GA operations manually. Since image comparison is an area of interest for online search providers and artificial intelligence, the algorithm may be automated in the near future.\\
\indent Looking at the future of electronics miniaturization, metal-induced nanowires on semiconductor surfaces are a promising route to investigate. The combination of the self-organizing nature of the nanowires and the templated nature of the semiconductor substrate results in a constant high quality of the devices. The scale of these devices pushes both theoretical and experimental capabilities to their limits, but also allows a more direct comparison between the two, leading to interesting new insights and a better understanding of the systems under study. Only through close collaboration between theoreticians and experimentalists will it be possible to fully understand such systems, and in the end be able to design nano-structures with predictable properties.

%Special thanks etc
%----------------------------------------------------------------------
\section{Acknowledgement}
\indent The author wishes to thank Arie van Houselt, Geert Brocks and Harold Zandvliet for the many interesting discussions on the nature of Pt-induced nanowires. He also wishes to thank Sylvia Wenmackers the interesting discussions on the nature of modeling and for proofreading the manuscript. Finally, the author wishes to dedicate this work to Mariette Galet, beloved grandmother, who passed away earlier this year.

%----------------------------------------------------------------------

%*************************************************************************************************************
% BIBLIOGRAPHY
%*************************************************************************************************************
% Put in \nocite{*} so all entries in the bibliography are included
%\nocite{*}
% This GATHER command is useful for when you want to use WinEdt's Gather functionality, i.e., type
% \cite{} and a popup box appears with all of your citations to choose from.  Leave the % on the next line.
% The commented way of writing Gather is the only correct way of doing it !!!
%%GATHER{danny.bib}
%%GATHER{notesTopRev.bib}
%\bibliography{danny,notesTopRev}

\begin{thebibliography}{207}
\bibitem[{fn:({\natexlab{a}})}]{fn:Moore}
\bibinfo{note}{In the original article, a doubling every year is calculated
  based on the five data-points available. Later, Moore refined the period to a
  doubling every two years, based on further developments in the industry.}
\bibitem[{fn:({\natexlab{c}})}]{fn:XCcorrgarbage}
\bibinfo{note}{In practice, this fourth term also includes a correction
  introduced by the assumption that the analytical form used for the kinetic
  energy term is exact.}
\bibitem[{fn:({\natexlab{d}})}]{fn:NWpt24nm}
\bibinfo{note}{Distances of $2.4$ nm (and more) are also observed, but these
  wires belong to different ``patches'' of NW-array on the same terrace.}
\bibitem[{fn:({\natexlab{e}})}]{fn:AuModel}
\bibinfo{note}{We will indicate the model classes with authors who first
  suggested such a model, irrespective of their either presenting an actual
  model in this class, or not. As such the more complex modifications of for
  example van Houselt's Giant Missing dimer Row model, as presented by Sauer,
  will also be indicated as HM.}
\bibitem[{fn:({\natexlab{g}})}]{fn:GGAenergy}
\bibinfo{note}{The energies presented are GGA energies, in general the LDA
  energies give a more unfavorable formation energy.}
\end{thebibliography}

\end{document}